\documentstyle[12pt]{article}

\def\R{ {\rm R \kern -.31cm I \kern .15cm}}
\def\C{ {\rm C \kern -.15cm \vrule width.5pt \kern .12cm}}
\def\Z{ {\rm Z \kern -.27cm \angle \kern .02cm}}
\def\N{ {\rm N \kern -.26cm \vrule width.4pt \kern .10cm}}
\def\1{{\rm 1\mskip-4.5mu l} }
\def\lsim{\raise0.3ex\hbox{$<$\kern-0.75em\raise-1.1ex\hbox{$\sim$}}}
\def\gsim{\raise0.3ex\hbox{$>$\kern-0.75em\raise-1.1ex\hbox{$\sim$}}}
\def\noi{\noindent}
\def\beq{\begin{equation}}   \def\eeq{\end{equation}}
\def\bea{\begin{eqnarray}}  \def\eea{\end{eqnarray}}
\def\nn{\nonumber}
\def\noi{\noindent}
\def\beeq{\begin{eqnarray}} \def\eeeq{\end{eqnarray}}
\newcommand\mysection{\setcounter{equation}{0}\section}
\renewcommand{\theequation}{\thesection.\arabic{equation}}
\newcounter{hran} \renewcommand{\thehran}{\thesection.\arabic{hran}}

\def\bmini{\setcounter{hran}{\value{equation}}
   \refstepcounter{hran}\setcounter{equation}{0}
   \renewcommand{\theequation}{\thehran\alph{equation}}\begin{eqnarray}}

\def\bminiG#1{\setcounter{hran}{\value{equation}}
\refstepcounter{hran}\setcounter{equation}{-1}
\renewcommand{\theequation}{\thehran\alph{equation}}
\refstepcounter{equation}\label{#1}\begin{eqnarray}}

\def\emini{\end{eqnarray}\relax\setcounter{equation}{\value{hran}}\renewcommand{\theequation}{\thesection.\arabic{equation}}}

\begin{document}
\centerline{\Large\bf Confinement with Kalb -- Ramond Fields}
\vskip 3 truecm

\centerline{\bf Ulrich ELLWANGER$^{\bf a}$, Nicol\'as WSCHEBOR$^{\bf
a,b}$}
\vskip 3 truemm
\centerline{$^{\bf a}$Laboratoire de Physique Th\'eorique\footnote{Unit\'e Mixte
de Recherche - CNRS - UMR 8627}}  
\centerline{Universit\'e de Paris XI, B\^atiment 210, F-91405 ORSAY
Cedex, France}
\vskip 3 truemm
\vskip 1 truecm

\centerline{$^{\bf b}$ Institutos de F\'{\i}sica} 
\centerline{Facultad de Ciencias (Calle Igu\'a 4225, esq. Mataojo)} 
\centerline{and Facultad de Ingenier\'{\i}a (C.C.30, CP 1100),
Montevideo, Uruguay}

\vskip 4 truecm

\begin{abstract}
We consider models with $N$ $U(1)$ gauge fields $A_{\mu}^n$, $N$ 
Kalb-Ramond fields $B_{\mu \nu}^n$, an arbitrary bare action and a
fixed UV cutoff $\Lambda$.  Under mild assumptions these can be
obtained as effective low energy theories of SU(N+1)  Yang Mills
theories in the maximal abelian gauge. For a large class of bare
actions they can be  solved in the large $N$ limit and exhibit
confinement. The confining phase is characterized  by an approximate
``low energy'' vector gauge symmetry under which the Kalb-Ramond
fields  $B_{\mu\nu}^n$ transform. The same symmetry allows for a
duality transformation showing that magnetic monopoles  have condensed.
The models allow for various mechanisms of confinement, depending on
which sources for $A_{\mu}^n$ or $B_{\mu \nu}^n$ are switched on, but
the area law for the Wilson loop  is obtained in any case.
\end{abstract}

\vskip 1.5 truecm
\noi LPT Orsay 01-69 \par
\noi June 2001 \par

\newpage
\pagestyle{plain}
\baselineskip 18pt

\mysection{Introduction}
\hspace*{\parindent} Monopole condensation is widely believed to be 
the mechanism responsible
for confinement in Yang-Mills (YM) theories \cite{1r,2r} (see 
\cite{3r} for a review). The
picture of a confining vacuum as a dual superconductor gives rise to 
models for the low energy
behaviour of YM theories which are based on (the dual of) the abelian 
Higgs model \cite{4r}.
Albeit quite successful phenomenologically \cite{5r,6r}, it is 
practically impossible to
derive these models in a systematic way from YM theories: By 
construction (i.e. the
assumption of duality) the gauge and scalar fields of the dual models 
are related non-locally
to the YM gauge fields, at least off-shell, hence local dual models 
can only capture the
semi-classical features of low energy YM theories. \par

Generalizing a formalism proposed by Julia and Toulouse \cite{7r} 
Quevedo and Trugenberger
\cite{8r} have presented quadratic actions for antisymmetric tensor 
fields, which describe the
condensation of various topological defects in various space-time 
dimensions  $d$. If applied to
monopole condensation in $d = 4$, this approach suggests a (local) 
action for a Kalb-Ramond
field $B_{\mu\nu}$ \cite{9r}. \par

There are two ways to understand the usefulness of 2-form fields 
$B_{\mu \nu}$ for the description of monopole condensation in $d = 4$:
First, the dual  abelian Higgs model in the broken phase necessarily
has to contain a pseudoscalar Goldstone  boson. Reversing the duality
transformation in $d = 4$, the dual of a pseudoscalar is a 2-form 
field $B_{\mu\nu}$. Second, in the presence of a monopole the abelian 
Bianchi identity $\partial_{[\mu} F_{\nu \rho]} = 0$ is violated  on
the world-line $x(t)$, and the  field strength $F_{\mu\nu}$ cannot be
written in the form $F_{\mu\nu} = \partial_{[\mu} A_{\nu]}$  in $x(t)$.
If monopoles have condensed in the vacuum, the abelian Bianchi identity
is  violated everywhere, and the field $F_{\mu\nu}$ is nowhere of the
form $\partial_{[\mu}A_{\nu]}$.  In this situation it is useful to
introduce an auxiliary field $B_{\mu\nu}$ for $F_{\mu\nu}$  \cite{10r}
in some analogy to the field strength formulation of YM theories
\cite{11r}  (it suffices, however, to introduce auxiliary fields
$B_{\mu\nu}$ for the components of  $F_{\mu\nu}$ associated to the $N_c
- 1$ $U(1)$ subgroups of $SU(N_c)$, see below). \par

All this does not imply that a 2-form field $B_{\mu\nu}$ appears in 
the physical spectrum of a theory with condensed monopoles: The
equations of motion for  $B_{\mu\nu}$, as derived from the full
effective action, can well be algebraic, i.e. the  propagator of
$B_{\mu\nu}$ has not necessarily a simple pole for finite negative
$p^2$ (in the  Minkowskian regime). The role of the $B_{\mu\nu}$
dependent terms in the full effective action is  then just to
parametrize in a compact way a subclass of higher derivative 
interactions among the fundamental gauge fields $A_{\mu}$, but the
modes in $B_{\mu\nu}$  which are not of the form $\partial_{[\mu}
A_{\nu]}$ correspond to topologically non-trivial gauge field
configurations $A_{\mu}$ associated to a non-vanishing monopole
density. \par

Monopole condensation is a purely abelian phenomenon even in the
context of YM theories. Abelian dominance becomes particularly clear
\cite{2ra} in the maximal abelian gauge (MAG) \cite{2rb}: the  $N_c^2 -
1$ gauge fields of $SU(N_c)$ are divided into the $N = N_c - 1$ $U(1)$
gauge fields  $A_{\mu}^n$, $n = 1 \dots N$, associated to the
generators of the $U(1)^N$ Heisenberg  subalgebra, and $N_c^2 - N_c$
``charged'' gauge fields $W_{\mu}^a$ (which carry at least one  of the
$N$ $U(1)$ charges) associated to the off-diagonal generators. The
continuum MAG  is characterized by the gauge condition ($D_{\mu}
W_{\mu})^a = 0$, where $D_{\mu}$ are the  $U(1)^N$-covariant
derivatives. Lattice results indicate that the $W_{\mu}^a$ gauge 
fields are massive in the MAG, with $M_W \sim 1$~GeV (for $SU(2)$)
\cite{12r}. A mechanism for  $W_{\mu}$ mass generation based on ghost
condensates has been proposed in \cite{13r}. \par

Combining the above considerations one is led to conclude that the 
natural low energy degrees of freedom of confining gauge theories are
$N$ $U(1)$ gauge  fields $A_{\mu}^n$, and $N$ Kalb-Ramond fields
$B_{\mu \nu}^n$ which serve to describe  monopole condensation. (Here
``degree of freedom'' does {\it not} necessarily signify an  asymptotic
physical state, but possibly just an auxiliary field with algebraic
equations of motion.) \par

The purpose of the present paper is the solution of a field theory 
with the corresponding field content, with a bare action as general as
possible, in the  large $N$ limit. The idea is to treat this theory as
an effective low energy theory valid below  some scale $\Lambda \sim
M_W$ (with, e.g., $\Lambda \sim 1$~GeV). Consequently we will 
supplement the model with an UV cutoff $\Lambda$, and allow for
arbitrary  non-renormalizable interactions $\sim \Lambda^{-p}$ (where
$p$ is determined by power counting) in the bare  action. The precise
numerical value of $\Lambda$ will not concern us here. \par

As we will find below, these models indeed describe confinement
provided the parameters of the bare action satisfy some inequality. In
the confining phase a duality transformation to an abelian Higgs model
can be specified (although the full dual action is still  non-local),
which shows that one describes indeed monopole condensation. Remarkably
the effective  action exhibits an approximate vector-like ``gauge
symmetry'' (a slight exaggeration,  since it is broken by derivative
terms) which allows to ``gauge away'' the low momentum  modes of the
$U(1)$ gauge fields $A_{\mu}^n$. They are swallowed by the fields
$B_{\mu\nu}^n$  in analogy with the Higgs-Kibble phenomenon with an
additional Lorentz index. This  symmetry of the low energy effective
action coincides with the one in \cite{9r} for the quadratic terms of
the action. Here, however, it holds for  part of the interactions (to
arbitrary powers in the fields) as well and, simultaneously, renders a
more general duality transformation possible. \par

The conventional criterium for confinement is the area law of the 
expectation value of the
Wilson loop. In an abelian theory the Stokes theorem allows to 
express the Wilson loop in
terms of a source localized on a ``Wilson surface'' bounded by the 
Wilson loop. In our
model such a source can be coupled either to $F_{\mu\nu}$ (denoted by 
$J^F$), to $B_{\mu
\nu}$ (denoted by $J^B$) or both of them. As we will discuss in 
detail, in the confining
phase of the model the area law is obtained in any case. We will 
show, on the other hand,
that only a particular linear combination of the sources $J^F$ and 
$J^B$ is consistent,
once the model is coupled to external quantum fields: Fluctuations 
of such quantum
fields would generate an infinite contribution to the action unless 
the long-range parts of
the correlators are cancelled. The remaining short-range parts of the 
correlators coincide
with various models for the expectation value of Yang-Mills field 
strength correlators
\cite{6r,14r}.  \par

A full-fledged computation of the string tension would require the 
solution of a coupled
set of equations of motion (in analogy to approaches based on the 
dual abelian Higgs model
\cite{6r}) which will not be attempted here. In any case the result will be
``non-universal'' in the sense that it depends on details of the bare 
action and the UV
cutoff. \par

In this paper we focus rather on the infrared regime of a confining 
gauge theory and the
remarkable fact that, given the above field content representing the 
feature of ``abelian
dominance'' and a non-perturbative computational scheme like the 
$1/N$ expansion, such
theories can actually be solved in spite of confinement. \par

The subsequent sections of the paper are organized as follows: In 
section 2 we will
discuss more precisely, how the model can be obtained from $SU(N_c)$ 
YM theories, specify
its field content and parametrize its bare action. Section 3 is 
devoted to its solution at
large $N$. Since the visibility of the confining phase necessitates a 
careful treatment of
the limit of an infrared cutoff $k^2$ tending to zero, we will first 
review the bosonic
$O(N)$ model at large $N$ in the broken phase: Already in this 
simple model a convex
effective potential, and hence a consistent treatment of the broken 
phase, requires the
introduction of $k^2 \not= 0$ and a discussion of non-analyticities 
for $k^2 \to 0$.
Subsequently the emergence of a confining phase of the $A_{\mu} - 
B_{\mu\nu}$ -- model is
seen to share similar non-analyticities with the $O(N)$ model, which 
originate again
from the convexity of the effective action (and {\it not} from the 
large $N$ limit).
\par

In section 4 we will discuss various properties of the confining 
phase: the possibility of a duality transformation, the emergence of a
vector-like approximate  gauge symmetry, the effective propagators of
the $A_{\mu} - B_{\mu\nu}$ -- system, the  ``response'' of the system
to external sources like the Wilson loop and the unsuccessful search
for  bound states (hence ``glueballs'' will only emerge once the
charged gluons of the $SU(N_c)$  YM theory is taken into account). In
section 5 we compare the large $N$ limit of our models  to the large
$N_c$ limit of $SU(N_c)$ YM (which turns out to be not the same),
discuss  generalizations of the bare action considered in section 3 as
well as the notion of universality, and conclude with an outlook. A 
short summary of the subsequent results has already appeared in
\cite{25r}.

\mysection{The A$_{\mu}$ - B$_{\mu \nu}$ -- model}
\hspace*{\parindent} Before giving a detailed description of our 
model, we will motivate it
starting with the pure $SU(N_c)$ YM theory in the MAG. As noted in the 
introduction, one splits
the gauge fields into the $N = N_c - 1$ $U(1)$ gauge fields 
$A_{\mu}^n$ and the charged gauge
fields $W_{\mu}^a$. The Euclidean partition functions reads

\beq
\label{2.1e}
{1 \over {\cal N}'} \int {\cal D} A\ {\cal D}W  {\cal D} c{\cal D} 
\bar{c} \  e^{-S_{YM} - S_{gf} - S_{gh}} \eeq

\noi where the YM action can be decomposed as

\beq
\label{2.2e}
S_{YM} = \int d^4x \left \{ F_{\mu \nu}^n F_{\mu \nu}^n + F_{\mu 
\nu}^a F_{\mu \nu}^a
\right \} \eeq

\noi and ${\cal N}'$ is a normalization. \par

The gauge fixing terms are of the form

\beq
\label{2.3e}
S_{gf} = \int d^4x \left \{ {1 \over 2 \alpha} \left ( \partial_{\mu} 
A_{\mu}^n \right )^2 + {1
\over 2 \alpha_W} \left ( D_{\mu} W_{\mu}^a \right )^2 \right \} \eeq

\noi with the $U(1)^N$-covariant derivative

\beq
\label{2.4e}
D_{\mu} W_{\mu}^a = \partial_{\mu} W_{\mu}^a + g \ f_{nb}^a \ 
A_{\mu}^n \ W_{\mu}^b \quad .
\eeq

\noi The form of the ghost interactions in $S_{gh}$ does not play any 
role subsequently. \par

Now we introduce Kalb-Ramond fields $B_{\mu\nu}^n$ as collective 
fields for the $N$ field strengths
$F_{\mu\nu}^n$ ({\it including} their non-abelian terms $\sim g \ 
f_{ab}^n\ W_{[\mu}^a W_{\nu ]}^b$) in the
partition function (\ref{2.1e}). This amounts to the multiplication 
of the integrand of (\ref{2.1e}) by

\beq
\label{2.5e}
1 = {1 \over {\cal N}_B} \int {\cal D} B \ e^{-S_B} \quad , \quad S_B 
= \int d^4x \left \{ {1 \over 4} \left (
F_{\mu\nu}^n - B_{\mu \nu}^n \right )^2 \right \} \quad .
  \eeq

\noi The new version of the partition function reads
\beq
\label{2.6e}
{1 \over {\cal N}} \int {\cal D} A \ {\cal D}B \ {\cal D}W {\cal D} c 
{\cal D} \bar{c} \ e^{-S_{YM} - S_{gf} - S_{gh} -
S_B}
  \eeq

\noi with ${\cal N} = {\cal N}' {\cal N}_B$. \par

Next, as stated in the introduction, we assume that the ${\cal D}W$ 
path integral is ``infrared save'': Lattice results show that the
$W_{\mu}^a$ gauge fields in the MAG  aquire a finite mass $M_W$, in
agreement with  the phenomenon of infrared abelian dominance in this
gauge  \cite{12r}. A mechanism for $W_{\mu}^a$ mass generation which
involves the four ghost interactions in $S_{gh}$ in  the MAG has been
proposed in \cite{13r}. The same mechanism (bi-ghost condensation)
renders the ``charged''  ghosts $c^a$, $\bar{c}^a$ massive. Now we
assume that the ${\cal D}W$ and ${\cal D}c{\cal D}\bar{c}$ path
integrals  in (\ref{2.6e}) are performed. However, in order not to
generate uncontrollable UV divergences, it  is advisable to integrate
simultaneously over the ``high momentum'' modes of $A_{\mu}^n$ and
$B_{\mu \nu}^n$:  These modes are needed in order to cancel UV
divergences which would violate the Slavnov-Taylor identities. \par

Let us thus split the measures ${\cal D}A$ and ${\cal D}B$ as

\beq
\label{2.7e}
{\cal D}A = {\cal D}A_{<M_W}\cdot {\cal D} A_{>M_W} \quad , \quad 
{\cal D}B = {\cal D} B_{<M_W} \cdot {\cal
D}B_{>M_W} \quad .
  \eeq

\noi The indices $<$$M_W$ and $>$$M_W$ denote modes with momenta in
the  corresponding ranges. Now we rewrite the partition function
(\ref{2.6e}) as

\beq
\label{2.8e}
{1 \over {\cal N}_<} \int {\cal D} A_{<M_W} \ {\cal D}B_{<M_W}\ 
e^{-S_{bare}(A,B,M_W)-S_{gf}}
  \eeq

\noi with

\beq
\label{2.9e}
e^{-S_{bare}(A,B,M_W)-S_{gf}} = {{{\cal N}_<} \over {\cal N}} \int
{\cal D} A_{>M_W}  \ {\cal D}B_{>M_W} \ {\cal D}W \ {\cal D}c{\cal
D}\bar{c} \ e^{-S_{YM} - S_{gf} - S_{gh} - S_b} \quad . \eeq

\noi $S_{gf}$ in (\ref{2.8e}) and on the left hand side of (\ref{2.9e})
corresponds to the first (abelian) term in (\ref{2.3e}). The partition
function (\ref{2.8e}) corresponds to the class of  models which are the
subject of the present paper. It represents the ``effective low energy
theory'' of  $SU(N_c)$ YM under the only assumption that
$S_{bare}(A,B,M_W)$ is local in the sense that an expansion in 
positive (but otherwise arbitrary) powers of derivatives exists, and
that the terms of lowest order (without  derivatives on $B_{\mu \nu}$
and $F_{\mu \nu}$) are non-zero. Its dependence on $M_W$ is then
dictated by  power counting.\par

Since the split of the path integral measures in 
(\ref{2.7e})--(\ref{2.9e}) is somewhat formal, we will very
briefly present an alternative definition of the model. To this end 
we introduce in the path integral
(\ref{2.6e}) a) sources $J_{A,\mu}^n$ and $J_{B, \mu \nu}^n$ for 
$A_{\mu}^n$ and $B_{\mu \nu}^n$, and b) an
infrared cutoff for $A_{\mu}^n$ and $B_{\mu \nu}^n$ in the form of a 
quadratic term $(-\Delta S_k(A,B))$ in
the exponent, with

\beq
\label{2.10e}
\Delta S_k (A,B) = {1 \over 2} \int {d^4p \over ( 2 \pi )^4} \left \{ 
A_{\mu}^n(p) \ R_{\mu\nu}^A(k,p) \
A_{\nu}^n (-p) + B_{\mu \nu}^n(p) \ R_{\mu \nu, \rho \sigma}^B (k,p) 
\ B_{\rho\sigma}^n(-p) \right \} \ .
  \eeq

\noi The tensors $R^A$ and $R^B$ should suppress low momentum modes, 
i.e. become large or even diverge for $p^2
< k^2$ (without having zero modes). First we choose $k = M_W$ and define

\beq
\label{2.11e}
e^{-W_{M_W}(J_A, J_B)} = {1 \over {\cal N}} \int {\cal D}A \ {\cal 
D}B \ {\cal D}W \ {\cal D} c{\cal D} \bar{c} \ e^{-
S_{YM} - \Delta S_{M_W} - S_{gf} - S_{gh} - S_B + J_A \cdot A + J_B 
\cdot B} \ .
  \eeq

\noi The Legendre transform of $W_{M_W}(J_A, J_B)$ with respect to 
$J_A$, $J_B$, the effective action
$\Gamma_{M_W}(A,B)$, plays the same role as $S_{bare}(A, B, M_W)$ in 
(\ref{2.8e}). \par

It is staightforward to replace $M_W$ in (\ref{2.11e}) by a varying 
IR cutoff $k$, take the derivative of both
sides of eq. (\ref{2.11e}) with respect to $k$, and derive an exact 
Wilsonian renormalization equation
\cite{15r} for $W_k (J_A, J_B)$ or its Legendre transform 
$\Gamma_k(A,B)$. The $A_{\mu} - B_{\mu\nu}$ -- model is
then {\it defined} by the corresponding Wilsonian RGE for $W_k$ or 
$\Gamma_k$, and the boundary condition at
$k = M_W$ given by (\ref{2.11e}). (The advantage of this procedure is 
that Slavnov-Taylor identities of the
underlying $SU(N_c)$ YM theory are well under control \cite{16r} even 
in the presence of cutoffs as in
(\ref{2.10e})). For convenience we will, however, most of the time 
refer to (\ref{2.8e}) as the definition of
the model. \par

Apart by locality, $S_{bare}(A,B,M_W)$ in (\ref{2.8e}) is constrained 
by the $N = N_c - 1\ U(1)$ gauge
symmetries preserved by the second term in $S_{gf}$ in (\ref{2.3e}), up 
to a covariant linear gauge fixing term
corresponding to the first term in (\ref{2.3e}) (this can easily be 
proven by deriving a corresponding Ward
identity). In the absence of fields carrying $U(1)$ charges 
$S_{bare}$ can thus depend on $A_{\mu}^n$ only
through its abelian field strengths $F_{\mu\nu}^n$, hence we write 
$S_{bare}(F,B,M_W)$. However, $S_{bare}$ is
not constrained by renormalizability; we have to allow for arbitrary 
powers in the fields and momenta.
Subsequently we assume that all dimensionful parameters are 
proportional to corresponding powers of $M_W$
(which also plays the role of an UV cutoff $\Lambda = M_W$, cf 
(\ref{2.8e})) and omit the corresponding powers of $M_W$.\par

Still, the most general dependence of $S_{bare}$ on $F$ and $B$ does 
not yet allow to solve the model in a
large $N$ limit. To this end one has to assume that $S_{bare}$ 
depends only on $O(N)$ invariant bilinear
operators (quadratic in $F$ or $B$). This corresponds to the 
assumption that the low energy effective action
of $SU(N_c)$ YM is dominated by quadratic $SU(N_c)$ invariants; then 
$O(N)$, under which $F_{\mu\nu}^n$ and
$B_{\mu\nu}^n$ transform as $N$-plets, appears as an accidental 
global symmetry. As we will show later the
qualitative features of our results do, however, not depend on the 
large $N$ limit and hence not on the
global $O(N)$ symmetry.  \par

Let us define the following three Lorentz scalar bilinear operators

\bea
\label{2.12e}
&&{\cal O}_1(x) = \sum_{n=1}^N F_{\mu\nu}^n(x) \ F_{\mu\nu}^n (x) 
\quad , \nn \\
&&{\cal O}_2(x) = \sum_{n=1}^N F_{\mu\nu}^n(x) \ B_{\mu\nu}^n (x) 
\quad , \nn \\
&&{\cal O}_3(x) = \sum_{n=1}^N B_{\mu\nu}^n(x) \ B_{\mu\nu}^n (x) \quad .
  \eea
 
One of the simplest bare actions of the present class of models, 
which exhibits already all of the essential
features, is given by

\beq
\label{2.13e}
S_{bare}(F,B) = \int d^4x \left \{ {\cal L}_{bare}({\cal O}_i) + {h 
\over 2} \left ( \partial_{\mu} 
\widetilde{B}_{\mu \nu}^n \right )^2 + {\sigma \over 2}  \left ( 
\partial_{\mu} B_{\mu \nu}^n \right )^2
\right \}
  \eeq

\noi with

\beq
\label{2.14e}
\widetilde{B}_{\mu \nu}^n = {1 \over 2} \ \varepsilon_{\mu 
\nu\rho\sigma} \ B_{\rho\sigma}^n \quad .
  \eeq

\noi ${\cal L}_{bare}$ in (\ref{2.13e}) may contain derivatives 
acting on the operators ${\cal O}_i$ defined in
(\ref{2.12e}). Somewhat ad hoc we choose just bilinear ``kinetic'' 
terms for $B_{\mu\nu}^n$; generalizations
are straightforward and will be discussed in section 5. An artefact 
of the ansatz (\ref{2.13e}) with constant
$h$, $\sigma$ will be the appearance of asymptotic states (poles in 
the propagators) with masses $\sim
M_W^2/h$; we do not expect the corresponding states to appear in an 
effective low energy theory of $SU(N_c)$
YM, since these would be $N$-plets of $O(N)$. They could easily be 
avoided by replacing $h$, $\sigma$ in
(\ref{2.13e}) by ``higher derivative'' terms

\beq
\label{2.15e}
h(p^2)\ , \ \sigma (p^2) \quad \hbox{with} \quad h(p^2) \ , \ 
\sigma(p^2) \to 0 \quad \hbox{for} \quad |p^2| \
\gsim\ M_W^2
  \eeq

\noi as is to be expected once these terms are generated by loops of 
the ``charged'' sector of $SU(N_c)$ YM.
Our subsequent main results will not depend on this choice, and unless 
stated otherwise we will continue to work
with constant $h$, $\sigma$ for simplicity. \par

In order to solve the model in the large $N$ limit we have to make 
assumptions on the $N$ dependence of the
parameters in $S_{bare}$. These assumptions can be summarized  by 
rewriting (\ref{2.13e}) as

\beq
\label{2.16e}
S_{bare}(F, B) = \int d^4x \left \{ N {\cal L'}_{bare} \left ( 
{{\cal O}_i \over N}\right ) + {h \over 2}
\left ( \partial_{\mu} \widetilde{B}_{\mu\nu}^n \right )^2 + 
{\sigma \over 2} \left ( \partial_{\mu} 
B_{\mu \nu}^n \right )^2 \right \}
  \eeq

\noi where the coefficients of ${\cal L'}_{bare}$ are independent of
$N$. \par

Our subsequent aim is the computation of the (Euclidean) generating 
functional $W(J)$ of connected Green
functions,

\beq
\label{2.17e}
e^{-W(J)} = {1 \over {\cal N}} \int {\cal D} A_{<M_W} \ {\cal 
D}B_{<M_W} \ e^{-S_{bare}(F,B) + \int d^4x \left
\{ {1 \over 2 \alpha} ( \partial_{\mu} A_{\mu}^n)^2 + J_{A,\mu}^n  
A_{\mu}^n + J_{B,\mu\nu}^n B_{\mu
\nu}^n \right \}}
  \eeq

\noi and the effective action obtained by the Legendre transform of 
$W(J)$. The indices $<$$M_W$ denote modes
with momenta $p^2 \ \lsim \ M_W^2$ as discussed below eq. 
(\ref{2.7e}). To this end we can add cutoff terms
$\Delta S(A,B)$ to the exponent in (\ref{2.17e}), which have 
been defined in (\ref{2.10e}). Now,
however, the cutoff functions $R^A$ and $R^B$ should suppress high 
momentum modes with $p^2 \ \gsim \ M_W^2$.

\mysection{The large N solution}

\hspace*{\parindent} The most convenient formalism for the solution  of
models of the form (\ref{2.16e}) in the large $N$ limit is the
introduction of auxiliary $O(N)$  singlet fields \cite{17r,18r}, which
we will use as well. Already in the case of bosonic $O(N)$ models with 
spontaneously broken symmetry there are, however, some subtleties
associated to the convexity of the effective  action: In order to
``see'' the convexity of the effective potential in the broken phase it
is  necessary to introduce an artificial infrared cutoff $k^2$ (in,
e.g., momentum space) and to study  carefully the limit $k^2 \to 0$.
Conventionally this phenomenon is investigated in the context of the 
Wilsonian Exact Renormalization Group \cite{19r}, but this formalism is
not obligatory. \par

In the case of models of the form (\ref{2.16e}) we observe a similar 
phenomenon: Now the ``confining
phase'' is observed only in the limit of an infrared cutoff $k^2$ 
tending to zero, and again its existence
is related to the convexity of the effective action. \par

In order to review these points, which play an essential role in our 
class of models, we will first discuss
them in the context of the bosonic $O(N)$ model in the broken phase 
in the next subsection 1. In the
subsection 2 we will return to the $A_{\mu} - B_{\mu \nu}$ model and 
use what we have learned before.

\subsection{The solution of the scalar O(N)-model}
\hspace*{\parindent}
The field content of this model is given by an $N$-plet of scalars 
$\varphi^n$ , $n = 1 \dots N$. It is
convenient to introduce a composite $O(N)$ singlet operator
 
\beq
\label{3.1e}
{\cal O}(x) = \varphi^n(x) \varphi^n(x) \quad .
\eeq

\noi The bare action of the model contains a bare potential, which 
can conveniently be expressed in terms of
the operator ${\cal O}(x)$. If one insists on renormalizability the 
bare potential contains only terms linear
and quadratic in ${\cal O}(x)$. Including its generic $N$ dependence 
it is then of the form

\beq
\label{3.2e}
V_{bare} \left ( {{\cal O} \over N} \right ) = {1 \over N}\int d^4x
\left \{  - {m^2 \over 2} \ {\cal O}(x) + {\lambda \over 4N} \ {\cal
O}(x)^2 \right \} \quad . \eeq

\noi (We chose a minus sign in front of $m^2$ in order to have $m^2$ 
positive in the broken phase.) In order
to implement both an infrared cutoff $k$ and an UV cutoff $\Lambda$ 
we supplement the kinetic terms with a
cutoff term $\Delta S_k^{\Lambda}$:

\beq
\label{3.3e}
S_{kin}(\varphi ) = {1 \over 2} \int d^4x \ \partial_{\mu}  
\varphi^n \partial_{\mu} \varphi^n + \Delta 
S_k^{\Lambda} \quad ,
  \eeq

\noi where

\beq
\label{3.4e}
\Delta S_k^{\Lambda} = {1 \over 2} \int {d^4p \over (2 \pi )^4} \ 
\varphi (p) \ R_k^{\Lambda}(p^2) \ \varphi
(-p) \quad .
  \eeq

\noi The cutoff function $R_k^{\Lambda}$ diverges for $p^2 \ \lsim \ 
k^2$ and $p^2 \ \gsim \ \Lambda^2$, and vanishes for \break $k^2 \
\lsim \ p^2 \ \lsim \ \Lambda^2$. The aim is to  compute the generating
functional $W_k(J)$ in the limit $k \to 0$, where

\beq
\label{3.5e}
e^{-W_k(J)} = {1 \over {\cal N}'} \int {\cal D} \varphi \ 
e^{-S_{kin}(\varphi ) - N  V_{bare} \left ( {{\cal O}
\over N} \right ) + \int d^4x \ J^n\varphi^n} \quad . \eeq

\noi The UV cutoff $\Lambda$ is considered to be fixed, and ${\cal 
N}'$ ensures $W_0(0) = 0$. The auxiliary
field method consists in writing

\beq
\label{3.6e}
e^{-NV_{bare}\left ( {{\cal O} \over N}\right )} = {1 \over {\cal 
N}_{\phi}} \int {\cal D} \phi \
e^{-NG_{bare}(\phi) - \int d^4x \ \phi \cdot {\cal O}} \quad .\eeq

\noi In the large $N$ limit the path integral in (\ref{3.6e}) is 
dominated by the stationary point(s), and the relation between
$V_{bare}$ and $G_{bare}$ becomes a Legendre transformation:

\beq
\label{3.7e}
N V_{bare} \left ( {{\cal O} \over N} \right ) = N 
G_{bare}(\widehat{\phi}) + \int d^4x \ \widehat{\phi}
\cdot {\cal O} \eeq

\noi where $\widehat{\phi} \equiv \widehat{\phi}({\cal O})$ solves

\beq
\label{3.8e}
\left [ {\delta \over \delta \phi (x)} N G_{bare}(\phi ) \right 
]_{\widehat{\phi}({\cal O})} = - \ {\cal O}(x)
\quad .  \eeq

\noi Conversely we have

\beq
\label{3.9e}
\left [ {\delta \over \delta {\cal O}(x)} N V_{bare} \left ( {{\cal O} 
\over N} \right ) \right ]_{
{\cal O}(\widehat{\phi})} = \widehat{\phi}(x) \quad . \eeq

\noi The correctness of the assignment of the powers of $N$ in 
(\ref{3.7e}) can be verified by rescaling
${\cal O} \to N  {\cal O}'$, after which each term is proportional to 
$N$ and the factor $N$ can be dropped.
\par

Generally, for given $V_{bare}$, one can solve (\ref{3.9e}) for 
${\cal O}(\widehat{\phi})$, insert it into
(\ref{3.7e}) and obtain $G_{bare}(\widehat{\phi})$. With $V_{bare}$ 
as in (\ref{3.2e}) one obtains

\beq
\label{3.10e}
G_{bare} (\phi ) = - {1 \over \lambda} \int d^4x \left ( \phi + {m^2 
\over 2} \right )^2 \quad ,
\eeq

\noi

\beq
\label{3.11e}
\widehat{\phi}({\cal O}) = - {m^2 \over 2} + {\lambda \over 2N} \ 
{\cal O}(x) \quad .
\eeq

\noi Note that $G_{bare}(\phi_i)$ is negative and unbounded from below
for $\phi_i \to \infty$, which is no point of concern as long as
$\phi$ is an auxiliary field with algebraic equations of motion. Using
(\ref{3.6e}) the expression (\ref{3.5e}) can be written as

\beq
\label{3.12e}
e^{-W_k(J)} = {1 \over {\cal N}} \int {\cal D} \phi \ 
e^{-NG_{bare}(\phi )} \int {\cal D} \varphi \ e^{-S_{kin}(\varphi ) -
\int d^4x (\phi \cdot {\cal O} - J^n\varphi^n)} \eeq

\noi with ${\cal N} = {\cal N}'{\cal N}_{\phi}$. \par

The ${\cal D}\varphi$ path integral is thus Gaussian and gives

\beq
\label{3.13e}
e^{-W_k(J)} = {1 \over {\cal N}} \int {\cal D} \phi \ 
e^{-NG_{bare}(\phi ) - N \Delta G(\phi ) + {1 \over 2}
\int d^4x_1 d^4x_2J^n(x_1) P(x_1,x_2, \phi )J^n(x_2)} \eeq

\noi with

\beq
\label{3.14e}
\Delta G(\phi ) = - {1 \over 2} \ Tr \log \left( P(x_1, x_2, \phi )
\right)\quad .
\eeq

\noi (A factor $N$ from the trace in (\ref{3.14e}) has already been 
included in (\ref{3.13e}).) The propagator
$P(x_1, x_2, \phi )$ has to be obtained from the quadratic terms in 
$\varphi$ in the exponent in
(\ref{3.12e}); it satisfies

\bea
\label{3.15e}
&&\int d^4 x_2 \left [ {\delta^2 S_{kin}(\varphi ) \over \delta 
\varphi^n(x_1) \ \delta \varphi^m(x_2)} + 2
\delta_{nm}\ \delta^4(x_1 - x_2) \phi (x_1) \right ]
 P^{mk}(x_2, x_3, \phi ) \nn \\ && = \delta_{nk} \ \delta^4 (x_1
- x_3) \eea

\noi and we always use $P^{mk} \equiv \delta_{mk}P$. \par

For constant $\phi$ its Fourier transform reads

\beq
\label{3.16e}
\widetilde{P}(p^2, \phi ) = {1 \over p^2 + R_k^{\Lambda}(p^2) + 2 \phi}
\eeq

\noi and (\ref{3.14e}) becomes

\beq
\label{3.17e}
\Delta G(\phi ) = {1 \over 2}  \int d^4x \int {d^4p \over (2 \pi )^4} 
\log \left ( p^2 + R_k^{\Lambda} (p^2)
+ 2 \phi \right ) \quad . \eeq

\noi For what follows it is most convenient to employ a ``sharp'' 
cutoff function $R_k^{\Lambda}(p^2)$, which
is infinite for $p^2 < k^2$ and $p^2 > \Lambda^2$ and zero for $k^2 < 
p^2 < \Lambda^2$. Up to irrelevant terms
independent of $\phi$ this gives the same result as corresponding 
cutoffs of the $d^4p$ integral, and one
obtains

\bea
\label{3.18e}
&&\Delta G(\phi ) = {1 \over 64 \pi^2} \int d^4x \Big \{ \left ( 
\Lambda^4 - 4 \phi^2 \right ) \log \left (
\Lambda^2 + 2 \phi \right )  \nn \\
&&\quad - \left ( k^2 - 4 \phi^2 \right ) \log \left ( k^2 + 2 \phi 
\right ) + 2
\phi \left ( \Lambda^2 - k^2 \right ) \Big \} \ . \eea

\noi Again, in the large $N$ limit the path integral (\ref{3.13e}) 
can be replaced by the stationary point,
and $W_k(J)$ becomes

\beq
\label{3.19e}
W_k(J) = N G_{bare}(\widehat{\phi}) + N \Delta G(\widehat{\phi}) - {1 
\over 2} \int d^4x_1d^4x_2 \{ J^n(x_1)
P(x_1,x_2, \widehat{\phi})J^n(x_2) \}\eeq

\noi where $\widehat{\phi} \equiv \widehat{\phi}(J)$ is the 
configuration which extremizes the right-hand side of (\ref{3.19e}).
First we look for the vacuum configuration  $\widehat{\phi}_0 \equiv
\widehat{\phi}(J=0)$ which has to solve, with $G_{bare}$ from
(\ref{3.10e}) and $\Delta G$  from (\ref{3.18e}),

\bea
\label{3.20e}
0 &=& \left [ {\delta \over \delta \phi} \left ( G_{bare}(\phi )
+ \Delta  G(\phi) \right ) \right ]_{\widehat{\phi}_0} \nn \\
&=& \left [ - {1 \over \lambda} (2 \phi + m^2) + {1 \over 16 \pi^2} 
\left ( 2 \phi \log \left ( {k^2 + 2 \phi \over \Lambda^2 + 2
\phi}\right )  + \Lambda^2 - k^2 \right )  \right ]_{\widehat{\phi}_0}
\ . \eea

\noi Clearly the solution $\widehat{\phi}_0$ of (\ref{3.20e}) depends
on the infra-red cutoff $k^2$. It is instructive to follow the solution
$\widehat{\phi}_0$, as  a function of the infrared cutoff $k^2$, from
$k^2 = \Lambda^2$ down to $k^2 \to 0$. For $k^2 = \Lambda^2$, where
$\Delta G(\phi)$ vanishes, we have

\beq
\label{3.21e}
\left . \widehat{\phi}_0 \right |_{k^2 = \Lambda^2} = - {m^2 \over 2}
\eeq

\noi which is negative by assumption and our convention in 
(\ref{3.2e}). For $k^2 < \Lambda^2$ the solution
$\widehat{\phi}_0$ has always to correspond to a positive argument of 
the logarithm in (\ref{3.20e}), which implies (assuming
$\Lambda^2 > m^2$, i.e. $\Lambda^2 + 2 \widehat{\phi}_0 > 0$)

\beq
\label{3.22e}
\left . \widehat{\phi}_0 \right |_{k^2} > - {k^2 \over 2}  \quad .
\eeq

\noi Under the additional assumption

\beq
\label{3.23e}
{\Lambda^2 \over 16 \pi^2} < {m^2 \over \lambda}
\eeq

\noi the following subtle behaviour is observed in the limit of 
vanishing infrared cutoff $k^2$: One has
simultaneously

\bea
\label{3.24e}
&&k^2 \to 0 \quad , \nn \\
&&\widehat{\phi}_0 \to 0_{-\varepsilon} \quad , \nn \\
&&{1 \over 8 \pi^2} \ \widehat{\phi}_0 \log \left ( {k^2 + 2 
\widehat{\phi}_0 \over \Lambda^2}\right ) \to {m^2 \over \lambda} -
{\Lambda^2 \over 16 \pi^2} = \hbox{finite} \quad . \eea

\noi Note that the non-analytic behaviour (\ref{3.24e}) could not be 
obtained if one would solve (\ref{3.20e}) right away for $k^2 = 0$. It
corresponds in fact to the flat ``inner  region'' of the effective
potential $V_{eff}(\varphi^n)$ in the broken phase, which is required
by the  convexity of the effective action (in terms of $\varphi^n$,
once $\phi$ has been eliminated by its equation of  motion). \par

In order to clarify this point we have to construct the $k$-dependent 
effective action $\Gamma_k(\varphi )$
from $W_k(J)$ by a Legendre transformation:

\beq
\label{3.25e}
\Gamma_k (\varphi ) = W_k (J) + \int d^4x \ J^n(x) \  \varphi^n(x)
\quad . \eeq

\noi Taking the functional derivative of (\ref{3.25e}) with respect to
$J^n(x)$, and using the expression (\ref{3.19e}) for $W_k(J)$, one
obtains for the ``classical'' field $\varphi^n$

\beq
\label{3.26e}
\varphi^n (x) = - {\delta W_k(J) \over \delta J^n(x)} = \int d^4x' \ 
P(x, x', \widehat{\phi}) \ J^n(x') \quad .  \eeq

\noi $\widehat{\phi}$ in (\ref{3.26e}) is the stationary point 
configuration of $W_k(J)$, as given by
(\ref{3.19e}), in the presence of sources:

\bea
\label{3.27e}
&& 0 = 
\left [
{\delta W_k(J) \over \delta \phi (x)} 
\right ]_{\widehat{\phi}} 
= \Big [ {\delta \over \delta
\phi (x)} \left ( N G_{bare}(\phi ) + N \Delta G(\phi ) \right ) 
\nn \\
&& + {1 \over 2} \int d^4x_1 \ d^4x'_1 \ d^4x_2 \ d^4x'_2 \ J^n(x_1) 
\ P(x_1, x'_1, \phi )\  {\delta P^{-1}
(x'_1, x'_2, \phi ) \over \delta \phi (x)}\nn \\
&&\cdot P(x'_2, x_2, \phi ) \ J^n (x_2) \Big ]_{\widehat{\phi}} \
. \eea

We have expressed the functional derivative of the term $\sim JPJ$ in 
$W_k(J)$ with respect to $\phi$ in terms
of $\delta P^{-1}/\delta \phi$ for two reasons: First, from 
(\ref{3.15e}), this functional derivative is
trivial:

\beq
\label{3.28e}
{\delta P^{-1}(x'_1, x'_2, \phi ) \over \delta \phi (x)} = 2 \delta^4 
(x'_1 - x) \delta^4 (x'_2 - x) \quad ;
\eeq

\noi Second, using (\ref{3.26e}) and (\ref{3.28e}), eq. (\ref{3.27e}) 
becomes very simple once written in
terms of $\varphi^n$: one finds

\beq
\label{3.29e}
0 = \left [ {\delta \over \delta \phi (x)} \left ( N G_{bare}(\phi  ) +
N \Delta G_{bare}(\phi ) \right ) + \varphi^n (x) \varphi^n (x) \right
]_{\widehat{\phi}} \quad . \eeq

\noi Eq. (\ref{3.29e}) allows to obtain $\widehat{\phi}$ as a 
functional of the classical fields
$\varphi^n(x)$ or ${\cal O}(x) = \varphi^n(x) \varphi^n(x)$. From 
(\ref{3.25e}) with (\ref{3.19e}) for $W(J)$,
the inverse of (\ref{3.26e}) for $J^n$ in terms of $\varphi^n$, and 
(\ref{3.15e}) for $P^{-1}$ one obtains for
the effective action $\Gamma_k(\varphi)$:

\beq
\label{3.30e}
\Gamma_k(\varphi ) = S_{kin}(\varphi ) + N \left ( 
G_{bare}(\widehat{\phi}) + \Delta G(\widehat{\phi}) \right
) + \int d^4x \ \widehat{\phi}(x) \cdot {\cal O}(x) \eeq

\noi where $\widehat{\phi}$ solves (\ref{3.29e}). For the effective 
potential for $k = 0$ we replace $\varphi^n$ by constants in
(\ref{3.30e}), i.e. we drop  $S_{kin}(\varphi )$ and write ${\cal O}(x)
\to {\cal O} \equiv \varphi^n \varphi^n$. Now we proceed to solve
(\ref{3.29e}). For constant ${\cal O}$ the solution
of (\ref{3.29e}) exhibits a similar behaviour in the limit of
vanishing  infrared cutoff as in (\ref{3.24e}): In the ``inner region''
of the effective potential, where

\beq
\label{3.31e}
  {\cal O} = |\varphi^n|^2 < {m^2 \over \lambda} - {\Lambda^2 \over 16
\pi^2}
\eeq

\noi (instead of (\ref{3.23e})), one obtains

\bea
\label{3.32e}
&&k^2 \to 0 \quad , \nn \\
&&\widehat{\phi} \to 0_{-\varepsilon} \quad , \nn \\
&&{1 \over 8{\pi}^2} \ \widehat{\phi} \ \log \left ( {k^2 + 2
\widehat{\phi} \over  \Lambda^2} \right ) \to {m^2 \over \lambda} -
{\Lambda^2 \over 16 \pi^2} - {\cal O} = \hbox{finite} \quad . \eea

\noi Hence, for $|\varphi^n|$ satisfying (\ref{3.31e}) we obtain
$\widehat{\phi} = 0$ in the expression (\ref{3.30e}) for the
effective potential, which is thus flat ($\varphi$-independent). \par

As mentioned above, the underlying reason for this phenomenon is not 
the large $N$ limit, but the convexity of
the effective action of $\varphi^n$ even in the phase with 
spontaneously broken symmetry. However, its
observation requires the introduction of an infrared cutoff $k^2$, 
and the study of the limits (\ref{3.32e}).
\par

We discussed this phenomenon in quite some detail, since its 
essential features are shared by the $A_{\mu} -
B_{\mu \nu}$ -- model, which is the subject of the next subsection.

\subsection{The solution of the A$_{\mu} - {\bf B}_{\mu}$ -- model}
\hspace*{\parindent} Now we use a very similar formalism in order to 
solve the model defined by the partition
function (\ref{2.17e}) in the large $N$ limit. As discussed below 
(\ref{2.17e}), we may implement both UV
cutoffs and infrared cutoffs $\Lambda$ and $k$, respectively, with 
the help of cutoff terms $\Delta S_k(A,B)$
as defined in (\ref{2.10e}) and appropriate properties of the cutoff 
functions $R^A$ and $R^B$. (Again we keep
the UV cutoff $\Lambda \sim M_W$ fixed, but study the limit $k \to 
0$.) Thus we write instead of (\ref{2.17e})

\beq
\label{3.33e}
e^{-W_k(J)} = {1 \over {\cal N}'} \int {\cal D} A \ {\cal D}B \ 
e^{-S_{bare}(F,B) - \Delta S_k(A,B) + \int
d^4x \left \{ {1 \over 2 \alpha} (\partial_{\mu}A_{\mu}^n)^2 + 
J_{A,\mu}^{n} A_{\mu}^n +
J_{B,\mu\nu}^{n}B_{\mu\nu}^n\right \} } \eeq

\noi with $S_{bare}(F,B)$ given by (\ref{2.16e}). \par

Next we introduce auxiliary fields $\phi_i$, one for each operator 
${\cal O}_i$ appearing in the argument of
${\cal L'}_{bare}$ in (\ref{2.16e}) (recall the definitions 
(\ref{2.12e}) for  these bilinear Lorentz scalar
operators). In analogy to eq. (\ref{3.6e}) we write

\beq
\label{3.34e}
e^{-N\int d^4x{\cal L'}_{bare}\left ( {{\cal O}_i \over N} \right )} = 
{1 \over {\cal N}_{\phi}} \int {\cal D} \phi_i \ e^{-NG_{bare}(\phi_i)
- \int d^4x\phi_i\cdot {\cal O}_i} \quad . \eeq

Again the path integral on the right-hand side of (\ref{3.34e}) can 
be replaced by its stationary point in the
large $N$ limit, and the relation between $G_{bare}$ and ${\cal 
L'}_{bare}$ becomes

\beq
\label{3.35e}
N \int d^4x\ {\cal L'}_{bare} \left ( {{\cal O}_i \over N}\right ) = N 
G_{bare}(\phi_i) + \int d^4x \ \phi_i \cdot
{\cal O}_i \quad . \eeq

In contrast to the scalar $O(N)$ model we do not assume here that 
${\cal L'}_{bare}$ contains only terms up to quadratic order in ${\cal
O}_i$. Hence the Legendre transform  (\ref{3.35e}) may be quite
involved in general, but this is not our point of concern: We may
equally define our  models right away through functionals
$G_{bare}(\phi_i)$. Note that we allow for derivative terms in ${\cal
L'}_{bare}$ where the derivatives act on ${\cal O}_i$; under  the
assumptions of locality specified below eq. (\ref{2.9e}) we will still
obtain a local expression for  $G_{bare}(\phi_i)$ including, of course,
derivatives acting on $\phi_i$. \par

Replacing (\ref{3.34e}) into (\ref{3.33e}) one finds (${\cal N} = 
{\cal N}' \cdot {\cal N}_{\phi}$)

\bea
\label{3.36e}
&&e^{-W_k(J)} = {1 \over {\cal N}} \int {\cal D} \phi_i \int {\cal 
D}A \ {\cal D}B \nn \\
&& \times e^{-NG_{bare}(\phi_i) - \Delta S_k(A,B) - \int d^4x \left \{ 
\phi_i\cdot {\cal O}_i + {h \over 2}
(\partial_{\mu}\tilde{B}_{\mu\nu}^n)^2 + {\sigma \over 2} 
(\partial_{\mu}B_{\mu\nu}^n)^2 + {1 \over 2 \alpha} (\partial_{\mu}
A_{\mu}^n)^2 - J_{A,\mu}^{n}A_{\mu}^n -  J_{B,\mu\nu}^{n} B_{\mu\nu}^n
\right \} } \nn \\ \eea

\noi and the ${\cal D}A{\cal D}B$ path integrals become again 
Gaussian. The result can formally be written as
follows: First we introduce the notation

\beq
\label{3.37e}
\varphi_r^n , \varphi_s^n = \left \{ A_{\mu}^n , B_{\mu \nu}^n \right 
\} \quad ,
\eeq

\noi i.e. the indices $r$, $s$, attached to the fields $\varphi^n$ 
denote both the different fields $A^n$, $B^n$
and the different Lorentz indices. Correspondingly we introduce 
sources $J_r^n$, $J_s^n$ for $J_{A,\mu}^n$ and
$J_{B,\mu\nu}^n$. Now (\ref{3.36e}) becomes

\beq
\label{3.38e}
e^{-W_k(J)} = {1 \over {\cal N}} \int {\cal D} \phi_i \ 
e^{-NG_{bare}(\phi_i) - N\Delta G(\phi_i) + {1 \over 2}
\int d^4x_1 d^4x_2 \left \{ J_r^n(x_1) P^{rs}(x_1, x_2, \phi_i) 
J_s^n(x_2) \right \} } \eeq

\noi with

\beq
\label{3.39e}
\Delta G(\phi_i) = - {1 \over 2} \ Tr \log \left ( P^{rs}(x_1, x_2, 
\phi_i ) \right ) \quad .
\eeq

Again the propagators $P^{rs}$ of the $A_{\mu}^n$, 
$B_{\mu\nu}^n$-system are, in principle, proportional to $\delta_{n,m}$
with $n, m = 1 \dots N$; we omitted these Kronecker  symbols for
simplicity and took care of the resulting contribution from the trace
in (\ref{3.39e}) by the  explicit factor $N$ multiplying $\Delta G$ in
(\ref{3.38e}). The propagators $P_{\mu,\nu}^{AA}$, 
$P_{\mu,\rho\sigma}^{AB}$ and $P_{\mu\nu,\rho \sigma}^{BB}$ depend on
the five terms $\Delta S_k$, $\phi_i \cdot {\cal  O}_i$, ${h \over 2}
(\partial \widetilde{B})^2$, ${\sigma \over 2} (\partial B)^2$ and ${1
\over 2} (\partial A)^2$ in the exponent of  (\ref{3.36e}). Simple
explicit expressions can be obtained only for constant fields $\phi_i$.
They are given, in  the Landau gauge $\alpha \to 0$, in the appendix.
For $\Delta G$ in (\ref{3.39e}) one obtains for constant $\phi_i$

\bea
\label{3.40e}
&&\Delta G(\phi_i) = {3 \over 2} \int d^4x \int {d^4p \over ( 2 \pi 
)^4} \Big [ \log \Big ( 
 \sigma p^2(\phi_1 + R^A(p^2) ) 
\nn \\
&&+ 4  ( \phi_1 + R^A(p^2) ) ( \phi_3 + R^B(p^2) ) 
- \phi_2^2  \Big ) 
+ \log \left ( hp^2 + 4 ( \phi_3 + R^B (p^2) ) \right 
) \Big ] \ . \nn \\ \eea

\noi The cutoff functions $R^A$ and $R^B$ parametrize the simplest 
form of $\Delta S_k$ given in (\ref{A.1e}) in
the appendix. Again we will employ, for simplicity, ``sharp'' cutoffs 
$R^A$, $R^B$ which are infinity for $p^2
< k^2$, $p^2 > \Lambda^2$ and zero for $k^2 < p^2 < \Lambda^2$. (We 
have checked that the stationary point in
$\phi_i$ described by eqs. (\ref{3.46e}) below appears also for 
smooth cutoff functions $R^A$, $R^B$.) Then
the $d^4p$ integral in (\ref{3.40e}) can be performed with the result

\bea
\label{3.41e}
&&\Delta G(\phi_i) = {3 \over 32\pi^2} \int d^4x \left [ \left ( 
\Lambda^4 - {(4\phi_1 \phi_3 - \phi_2^2)^2
\over \sigma^2 \phi_1^2} \right ) \log \left ( \Lambda^2 \sigma 
\phi_1 + 4 \phi_1 \phi_3 - \phi_2^2 \right )
\right . \nn \\
&&- \left ( k^4 - {(4 \phi_1 \phi_3 - \phi_2^2 )^2 \over \sigma^2 
\phi_1^2} \right ) \log \left ( k^2 \sigma
\phi_1 + 4 \phi_1 \phi_3 - \phi_2^2 \right ) + {4 \phi_1 \phi_3 - 
\phi_2^2 \over \sigma \phi_1} \left (
\Lambda^2 - k^2 \right ) \nn \\
&& +\left ( \Lambda^4 - {16 \phi_3^2 \over h^2} \right ) \log  \left (
\Lambda^2 h + 4 \phi_3 \right ) - \left ( k^4 - {16 \phi_3^2 \over
h^2}\right ) \log \left ( k^2 h +  4\phi_3 \right ) \nn \\
&&\left . + 4{\phi_3 \over h} \left ( \Lambda^2 - k^2 \right ) \right ]
  .   \nn \\
\eea

\noi This expression for $\Delta G(\phi_i)$ has to be inserted into 
(\ref{3.38e}), and again the ${\cal
D}\phi_i$ path integral is dominated by its stationary point(s). 
Hence $W_k(J)$ becomes

\beq
\label{3.42e}
W_k(J) = N G(\widehat{\phi}_i) - {1 \over 2} \int d^4x_1 \ d^4x_2 
\left \{ J_r^n(x_1) P^{rs}(x_1, x_2,
\widehat{\phi}_i)J_s^n (x_2) \right \} \eeq

\noi where

\beq
\label{3.43e}
G(\phi_i) = G_{bare} (\phi_i) + \Delta G(\phi_i)
\eeq

\noi and $\widehat{\phi}_i \equiv \widehat{\phi}_i(J)$ satisfy the 
three equations (recall $i = 1,2,3$)

\beq
\label{3.44e}
\left [ {\delta \over \delta \phi_i} \Big (NG(\phi_i) - {1 \over 2}
\int  d^4x_1 \ d^4x_2 \ J_r^n(x_1) P^{rs}(x_1, x_2, \phi_i) J_s^n
(x_2) \Big ) \right ]_{\widehat{\phi}_i(J)} = 0 \ . \eeq

\noi First we are interested in the vacuum configurations 
$\widehat{\phi}_i^0 \equiv \widehat{\phi}_i(J=0)$.
The three stationary point equations (\ref{3.44e}) with $\Delta G$ 
given by (\ref{3.41e}) simplify considerably
if we switch from the three independent fields $\phi_1$, $\phi_2$, 
$\phi_3$ to $\phi_1$, $\phi_3$, $\Sigma$
with

\beq
\label{3.45e}
\Sigma = {4 \phi_1 \phi_3 - \phi_2^2 \over \sigma \phi_1} \quad .
\eeq

\noi Then the stationary point equations for $J = 0$ become (a factor 
$N$ can be dropped)

\bminiG{3.46e}
\label{3.46ae}
\left [ {\delta G_{bare} \over \delta \phi_1} + {3 \over 32 \pi^2 
\phi_1} \left ( \Lambda^4 - k^4 \right )
\right ]_{\widehat{\phi}_i^0} = 0 \ , \eeeq  \beeq
  \label{3.46be}
\left [ {\delta G_{bare} \over \delta \phi_3} + {3 \over 4 \pi^2 h^2} 
\left \{ 4 \phi_3 \log \left (
{4 \phi_3 + hk^2 \over 4 \phi_3 + h \Lambda^2} \right ) + h \left ( 
\Lambda^2 - k^2 \right ) \right \} \right
]_{\widehat{\phi}_i^0} = 0 \ , \eeeq
\beeq
  \label{3.46ce}
\left [ {\delta G_{bare} \over \delta \Sigma} + {3 \over 16 \pi^2 } 
\left \{ \Sigma \log \left (
{\Sigma + k^2 \over \Sigma + \Lambda^2} \right ) +  
\Lambda^2 - k^2 \right \} \right
]_{\widehat{\phi}_i^0} = 0 \ .
\emini

\noi Clearly eqs. (\ref{3.46e}) can not be solved without knowledge 
of $G_{bare}$ (or its Legendre transform
${\cal L}_{bare}$, cf. (\ref{3.35e})), and they may have several 
solutions. In the latter case the one with the
lowest vacuum energy has to be chosen. This issue, together with the 
physical interpretation of the auxiliary
fields $\phi_i$, will be discussed below. \par

Let us now assume that, in some analogy to (\ref{3.23e}) in the case of
the scalar $O(N)$ model,

\beq
\label{3.47e}
\left. {\delta G_{bare} \over \delta \Sigma} \right
|_{\widehat{\Sigma}^0 = 0}  +{3 \over 16 \pi^2} \ \Lambda^2  < 0
\quad . \eeq

\noi Then, provided $\widehat{\Sigma}^0 (k^2 = \Lambda^2) < 0$, the
solution of (\ref{3.46ce}) behaves as follows in the limit of vanishing
infra-red cutoff: We have simultaneously 

\bminiG{3.48e}
\label{3.48ae}
k^2 \to 0 \quad ,
\eeeq  \beeq
  \label{3.48be}
\widehat{\Sigma}^0(k^2) \to 0_{-\varepsilon} \quad ,
\eeeq
\beeq
\label{3.48ce}
\left . {3 \over 16 \pi^2} \widehat{\Sigma}^0 \log \left (
{\widehat{\Sigma}^0 + k^2 \over  \widehat{\Sigma}^0 + \Lambda^2}\right )
\to  - \ {\delta G_{bare} \over \delta \Sigma} \right
|_{\widehat{\Sigma}^0 = 0}  -{3 \over 16 \pi^2} \ \Lambda^2 
\ . \emini

\noi From (\ref{3.45e}) the stationary  point (\ref{3.48be}), if it
exists, corresponds to (assuming $\widehat{\phi}_1^0 < \infty$)

\beq
\label{3.49e}
4 \widehat{\phi}_1^0 \ \widehat{\phi}_3^0 - \widehat{\phi}_2^{0^2} = 0
\quad .
\eeq

\noi Eqs. (\ref{3.46ae}) and (\ref{3.46be}) represent two additional 
constraints on the three fields
$\widehat{\phi}_1^0$, $\widehat{\phi}_2^0$ and $\widehat{\phi}_3^0$, 
which are thus all determined. As in the
scalar $O(N)$ model this stationary point would not have been 
observed if one puts $k^2 = 0$ from the start.
\par

Subsequently we will denote the stationary point (\ref{3.48e}), where 
(\ref{3.49e}) holds, the ``confining phase'' of the model. As stated
above its existence depends on the validity of (\ref{3.47e}); clearly,
this condition requires no ``fine-tuning'' of $G_{bare}$
or ${\cal L}_{bare}$. \par

Next we construct the effective action of the model. Of course it is 
given by the Legendre transform of
$W_k(J)$ in (\ref{3.42e}):

\beq
\label{3.50e}
\Gamma_k (A,B) = W_k(J) + \int d^4x \left ( J_{A, \mu}^n A_{\mu}^n 
+ J_{B, \mu \nu}^n B_{\mu \nu}^n \right
) \quad . \eeq

\noi Following the steps discussed from eqs. (\ref{3.26e}) to 
(\ref{3.30e}) in the context of the bosonic
$O(N)$ model one obtains

\beq
\label{3.51e}
\Gamma_k(A,B) = \Delta S_k + N G(\widehat{\phi}_i) + \int d^4x \left 
( \widehat{\phi}_i {\cal O}_i + {h \over
2} \left ( \partial_{\mu} \widetilde{B}_{\mu\nu}^n \right )^2 + 
{\sigma \over 2} \left ( \partial_{\mu}
B_{\mu\nu}^n \right )^2 \right ) \eeq

\noi with $G(\widehat{\phi}_i)$ as in (\ref{3.43e}), and where it is 
again convenient to replace the sources in the stationary point
equations (\ref{3.44e}) by the fields $A_{\mu}^n$,   $B_{\mu\nu}^n$ or
the corresponding operators ${\cal O}_i$ (\ref{2.12e}):

\beq
\label{3.52e}
\left [ {\delta \over \delta \phi_i} N G(\phi_i) + {\cal O}_i 
\right ]_{\widehat{\phi}_i(A,B)} = 
\left [\ {\delta \Gamma_{k} \over \delta \phi_i}\ 
\right]_{\widehat{\phi}_i(A,B)} = 0 \quad .
\eeq

\noi In the vacuum we require $A_{\mu}^n$, $B_{\mu\nu}^n = 0$ and 
hence ${\cal O}_i = 0$ in order not to
break Lorentz invariance, and then (\ref{3.52e}) coincides with 
(\ref{3.44e}) for $J = 0$. \par

Let us briefly discuss the relation between the condition 
$\widehat{\Sigma}^0(k^2 \to \Lambda^2) < 0$ for the
confining phase to appear, and the convexity of the effective action 
in the case of a non-convex bare
action. To this end we consider constant configurations 
$F_{\mu\nu}^n$, $B_{\mu\nu}^n$ in the expression
(\ref{3.51e}) for $\Gamma_k(A, B)$. \par

Given the dependence of ${\cal O}_i$ on $F_{\mu\nu}^n$, 
$B_{\mu\nu}^n$ in (\ref{2.12e}) it is straightforward
to compute the determinant of second derivatives of $\Gamma_k({\cal 
O}_i)$ with respect to $F_{\mu\nu}^n$,
$B_{\mu\nu}^n$. If one drops $\Delta S_k$ in (\ref{3.51e}), this 
determinant is negative if

\beq
\label{3.53e}
4 \widehat{\phi}_1 \ \widehat{\phi}_3 - \widehat{\phi}_2^2 < 0 \quad .
\eeq

\noi In this case $\Gamma_k$ without $\Delta S_k$ would be non-convex 
around the origin $F = B = 0$ in field
space. Under the assumption $\widehat{\phi}_1 > 0$ (otherwise the 
term $\sim F^2$ in $\Gamma_k$ would have the
``wrong'' sign) (\ref{3.53e}) corresponds to the negativity of 
$\widehat{\Sigma}$, cf. eq. (\ref{3.45e}). The
confining stationary point $\widehat{\Sigma} \to 0_{-\varepsilon}$ 
for $k^2 \to 0$ corresponds thus to the
``ironing'' of a non-convex bare action as in the case of a 
non-convex bare potential in the bosonic $O(N)$
model in the broken phase. \par

The condition for the reach of the confining phase for $k^2 \to 0$ in 
the presence of non-vanishing
backgrounds $A_{\mu}^n$, $B_{\mu\nu}^n$ and hence ${\cal O}_i$ 
differs somewhat from the positivity of the
right-hand side of (\ref{3.48ce}). After switching to the independent 
variables $\phi_1$, $\phi_3$ and
$\Sigma$, the three equations (\ref{3.52e}) can be written as

\bminiG{3.54e}
\label{3.54ae}
\left [ {\delta \over \delta \phi_1} NG(\phi_i) + {\cal O}_1 + 
\sqrt{{4\phi_3 - \sigma \Sigma \over 4 \phi_1}} \
{\cal O}_2 \right ]_{\widehat{\phi}_i} = 0 \ ,\eeeq  \beeq
  \label{3.54be}
\left [ {\delta \over \delta \phi_3} NG(\phi_i) + {\cal O}_3 + 
\sqrt{{4\phi_1\over 4 \phi_3 -
\sigma \Sigma }} \ {\cal O}_2 \right ]_{\widehat{\phi}_i} = 0 \ ,
\eeeq
\beeq
  \label{3.54ce}
\left [ {\delta \over \delta \Sigma} NG(\phi_i) - {\sigma \over 2}  
\sqrt{{\phi_1\over 4 \phi_3 -
\sigma \Sigma }} \ {\cal O}_2 \right ]_{\widehat{\phi}_i} = 0 \ .
\emini

\noi In principle we can solve (\ref{3.54ae}) and (\ref{3.54be}) for 
$\widehat{\phi}_1({\cal O}_i, \widehat{\Sigma})$ and
$\widehat{\phi}_3({\cal  O}_i,\widehat{\Sigma})$. Then equation
(\ref{3.54ce}) has the ``confining'' solution $\widehat{\Sigma}(k^2)
\to O_{-\varepsilon}$ for $k^2 \to 0$ if

\beq
\label{3.55e}
\left [ N \left \{ {\delta G_{bare}(\widehat{\phi}_i) \over \delta 
\widehat{\Sigma}} + {3 \over 16 \pi^2}
\Lambda^2 \right \} - {\sigma \widehat{\phi}_1 \over 2 
\widehat{\phi}_2} {\cal O}_2 \right ]_{\widehat{\Sigma}
= 0} < 0 \quad . \eeq

\noi Notably we then have, for $k^2 \to 0$,

\beq
\label{3.56e}
4 \widehat{\phi}_1 \widehat{\phi}_3 - \widehat{\phi}_2^2 = 0
\eeq

\noi even for non-vanishing classical fields $A_{\mu}^n$, 
$B_{\mu\nu}^n$ or ${\cal O}_i$. Generally, once the
right-hand side of (\ref{3.48ce}) is positive, there exists a finite 
region in the space of operators ${\cal
O}_i$ where (\ref{3.55e}) holds as well. \par

Many remarkable properties of the effective action (\ref{3.51e}) in 
the confining phase (\ref{3.56e}) will be
discussed in the next chapter. To close the present section, we will 
clarify the physical meaning of the
auxiliary fields $\phi_i$ which parametrize the vacua of the model. \par

To this end we return to the partition function (\ref{3.36e}). 
Instead of ``switching on'' sources $J$ for the
fields $A_{\mu}^n$, $B_{\mu\nu}^n$ we will consider sources $K_i(x)$ 
for the composite operators ${\cal O}_i$
and define

\bea
\label{3.57e}
&&e^{-W_k(K)} = {1 \over {\cal N}} \int {\cal D} \phi_i \int {\cal D} 
A {\cal D} B \nn \\
&& e^{-NG_{bare} (\phi_i) - \Delta S_k (A, B) - \int d^4x \left \{ 
\phi_i {\cal O}_i + {h \over 2} \left (
\partial_{\mu} \widetilde{B}_{\mu\nu}^n \right )^2 + {\sigma \over 2} 
\left ( \partial_{\mu} B_{\mu\nu}^n
\right )^2 + {1 \over 2 \alpha} \left ( \partial_{\mu} 
A_{\mu}^n\right )^2 - K_i {\cal O}_i \right \} } \ . \nn \\ &&
\eea

\noi (We could have kept the sources $J$; we have omitted them just 
for simplicity.) Evidently $W_k(K)$
generates the Green functions of the composite operators ${\cal 
O}_i$; notably we have

\beq \label{3.58e}
<{\cal O}_i(x) >_K = - {\delta W_k(K) \over \delta K_i(x)} \quad .
\eeq

\noi One observes in (\ref{3.57e}) that the sources $K_i$ couple to 
the bilinear operators ${\cal O}_i$ in the
same way as the auxiliary fields $\phi_i$. Repeating the steps which 
have led to eq. (\ref{3.42e}) for
$W_k(J)$ one thus obtains immediately

\beq
\label{3.59e}
W_k(K) = N G_{bare} (\widehat{\phi}_i) + N \Delta G\left ( 
\widehat{\phi}_i - K_i \right )
\eeq

\noi with $\Delta G$ as in (\ref{3.41e}), and where the fields 
$\widehat{\phi}_i \equiv \widehat{\phi}_i(K)$
satisfy

\beq
\label{3.60e}
\left [ {\delta \over \delta \phi_i} \Big (G_{bare}(\phi_i) + \Delta
G\left  ( \phi_i - K_i \right ) \Big ) \right ]_{\widehat{\phi}_i(K)} =
0 \quad . \eeq

\noi Using (\ref{3.59e}) and (\ref{3.60e}) in (\ref{3.58e}) one finds

\beq
\label{3.61e}
<{\cal O}_i(x)>_K \ = N \ {\delta \Delta G(\widehat{\phi}_i - K_i ) 
\over \delta \widehat{\phi}_i (x)} \quad .
\eeq

\noi Hence in the classical limit (or for an infrared cutoff $k^2 \to 
\Lambda^2$) where $\Delta G$ vanishes,
the vacuum expectation values of the composite operators ${\cal O}_i$ 
vanish as they should. \par

In the limit of vanishing source $K_i$ we can use (\ref{3.60e}) again
to write
 
\beq
\label{3.62e}
<{\cal O}_i(x)>_{K=0} \ = - N \ {\delta G_{bare}(\widehat{\phi}_i) 
\over \delta \widehat{\phi}_i}
\eeq

\noi where the auxiliary fields $\widehat{\phi}_i$ satisfy the 
stationary point equation (\ref{3.60e}) with $K_i = 0$. At first sight 
(\ref{3.62e}) seems to be a trivial consequence of the Legendre
transform (\ref{3.35e}), but now the  $\widehat{\phi}_i$ are the
extrema of $G_{bare} + \Delta G$ including the quantum contribution. In
any case  (\ref{3.62e}) shows that different vevs $\widehat{\phi}_i$
parametrize different vevs $<{\cal O}_i>$ (recall  their definitions
(\ref{2.12e})). This relation would be linear if $G_{bare}$ would be
quadratic in $\phi_i$ (as in the renormalizable $O(N)$ model). \par

Since the $\phi_i$ are auxiliary fields for composite operators the 
interpretation of the effective action
(\ref{3.51e}) (at $k = 0$, where $\Delta S_k$ vanishes, and for 
constant field configurations) as an energy
density is not straightforward \cite{18r}, if the $\phi_i$ are 
considered as independent variables. Notably
$\Gamma (\phi_i, A, B)$ is generically unbounded from below for 
$|\phi_i| \to \infty$ due to the negativity of
$G_{bare}(\phi_i)$. (Such a feature of a potential $V$ involving 
auxiliary fields is well-known from
supersymmetric theories, once auxi\-liary $F$- or $D$-fields are 
introduced in order to complete the
supermultiplets. Then one has e.g., $V(F, \varphi ) = - |F_i|^2 + 
F_i W_i + h.c.$ where $W_i$ is the
derivative of the superpotential $W(\varphi_i)$). At the stationary 
points $\phi_i = \widehat{\phi}_i$,
however, the interpretation of $\Gamma (\widehat{\phi}_i, A=B=0)$ as 
an energy density can be maintained
\cite{18r} and the energy densities of different vacua can be compared. \par

The interesting properties of the effective action in the confining 
phase with (\ref{3.49e}) will be the
subject of the next section.

\mysection{Properties of the confining phase}
\hspace*{\parindent} In the following we assume that the necessary 
condition (\ref{3.55e}) for the existence
of a confining phase is satisfied within a finite regime of field 
configurations $A_{\mu}^n$, $B_{\mu\nu}^n$
(and corresponding bilinears ${\cal O}_i$), and that it represents the 
lowest (or only) vacuum of the
model. For $k^2 \to 0$ the effective action (\ref{3.51e}) reads

\beq
\label{4.1e}
\Gamma (A,B) = N G (\widehat{\phi}_i) + \int d^4x \left \{ 
\widehat{\phi}_i {\cal O}_i + {h \over 2} \left (
\partial_{\mu} \widetilde{B}_{\mu\nu}^n \right )^2 + {\sigma \over 2} 
\left ( \partial_{\mu} B_{\mu\nu}^n \right
)^2 \right \} \eeq

\noi with $G(\phi_i)$ as in (\ref{3.43e}), and the $\widehat{\phi}_i$ 
satisfy (\ref{3.52e}). The confining
phase is characterized by the validity of eq. (\ref{3.56e}), $4 
\widehat{\phi}_1 \widehat{\phi}_3 -
\widehat{\phi}_2^2 = 0$. \par

First we consider a new kind of $N$ $U(1)$ gauge transformations 
which involve vector-like gauge parameters
$\Lambda_{\mu}^n$ \cite{9r}:

\bea
\label{4.2e}
&&\delta A_{\mu}^n(x) = \Lambda_{\mu}^n(x) \ , \ \delta 
F_{\mu\nu}^n(x) = \partial_{\mu} \Lambda_{\nu}^n
(x) - \partial_{\nu} \Lambda_{\mu}^n(x) \equiv \Lambda_{\mu\nu}^n 
(x) \ , \nn \\
&&\delta B_{\mu}^n(x) = \sqrt{{\widehat{\phi}_1 \over 
\widehat{\phi}_3}} \ \Lambda_{\mu\nu}^n (x) \quad
.\eea

\noi Using eq. (\ref{3.56e}) in order to eliminate 
$\widehat{\phi}_2$, (choosing $\widehat{\phi}_2$ positive,
otherwise the sign in $\delta B$ has to be changed) and (\ref{2.12e}) 
for the definitions of the bilinear
${\cal O}_i$, one finds that the terms $\widehat{\phi}_i {\cal O}_i$ 
in (\ref{4.1e}) can be written as

\beq
\label{4.3e}
\widehat{\phi}_i {\cal O}_i = \sum_n \left ( 
\sqrt{\widehat{\phi}_1} \ F_{\mu \nu}^n +
\sqrt{\widehat{\phi}_3} \ B_{\mu \nu}^n \right )^2 \quad . \eeq

\noi Hence these terms are invariant under the gauge transformations 
(\ref{4.2e}). Next we have a look at the
term $\sim (\partial_{\mu} \widetilde{B}_{\mu\nu}^n)^2$. Thanks to a 
Bianchi identity one obtains

\beq
\label{4.4e}
\delta \left ( \partial_{\mu} \widetilde{B}_{\mu\nu}^n \right )^2 = 
0 + {\cal O} \left ( \partial_{\mu} 
\widehat{\phi}_i \right ) \quad ,  \eeq

\noi i.e. it is also invariant up to derivatives acting on 
$\widehat{\phi}_i$ which appear in the variation
$\delta B_{\mu\nu}^n$ in (\ref{4.2e}). Recall that the 
$\widehat{\phi}_i$ are fixed in terms of the bilinear
operators ${\cal O}_i$ by (\ref{3.52e}); for constant configurations 
${\cal O}_i$ the $\widehat{\phi}_i$
would also be constant, and the symmetry breaking contribution to 
(\ref{4.4e}) is thus proportional to
$\partial_{\mu}{\cal O}_i$. \par

The last term $\sim (\partial_{\mu} B_{\mu\nu}^n)^2$ in (\ref{4.1e}) 
behaves as a gauge fixing term of the
symmetry (\ref{4.2e}), and altogether one obtains a Ward identity

\beq
\label{4.5e}
\delta \Gamma (A,B) = - \sqrt{{\widehat{\phi}_1 \over 
\widehat{\phi}_3}} \ \sigma \left ( \partial_{\mu}
 B_{\mu\nu}^n \right )  \left ( \partial_{\mu}
 \Lambda_{\mu\nu}^n \right ) + {\cal O} \left ( \partial_{\mu}  
{\cal O}_i \cdot \left ( \partial_{\mu} B
\right )^2 \right ) \ .  \eeq

It is natural to expect that the symmetry (\ref{4.2e}) is broken by 
higher derivative terms (beyond the gauge
fixing term): The bare action $S_{bare}$ (\ref{2.16e}) of the model 
does certainly not exhibit the symmetry
(\ref{4.2e}) (or satisfy the Ward identity (\ref{4.5e})), and the 
Green functions at large non-exceptional
Euclidean momenta with $p^2 \to \Lambda^2$ are generated by 
$S_{bare}$. This fact is realized by the
dependence of the effective action on higher derivative terms. The 
symmetry (\ref{4.2e}) is thus a pure ``low
energy'' phenomenon. \par

Note, however, that the ``violation'' of the Ward identity 
(\ref{4.5e}) is of higher order in the fields, and
is proportional to derivatives acting on the bilinears ${\cal O}_i$. 
Hence the symmetry (\ref{4.2e}) is
realized on modes $A_{\mu}^n$ and $B_{\mu\nu}^n$ which may be rapidly 
oscillating, but correspond to constant
bilinear configurations ${\cal O}_i$. (These modes are associated 
with Green functions with ``exceptional''
external momenta.)  \par

The implication of the gauge symmetry (\ref{4.2e}) on these modes of 
the $U(1)$ gauge fields $A_{\mu}^n$ is
evidently that they can be ``gauged away'' and ``eaten'' by the 
(massive or even infinitely massive)
Kalb-Ramond fields $B_{\mu\nu}^n$ \cite{9r}.\par

Correspondingly the (approximate) gauge symmetry (\ref{4.2e}) will 
have consequences on the ``response'' of
the model with respect to couplings to external sources $J^A$ and 
$J^B$. Before studying this issue we will
discuss the behaviour of the effective action (\ref{4.1e}) under a 
duality transformation, whose existence is closely
related to the symmetry (\ref{4.2e}). \par

Let us recall the essential features of a duality transformation: It 
corresponds to the introduction of
dual fields such that the Bianchi identities associated to the dual 
fields correspond to the equations of
motion of the ``original'' fields (up to gauge fixing 
terms), and the Bianchi identities associated to
the ``original'' fields to the equations of motion of the dual fields. \par

Using (\ref{4.3e}), the equations of motion for $A_{\mu}^n$ and 
$B_{\mu\nu}^n$ are

\bminiG{4.6e}
\label{4.6ae}
\partial_{\mu} \left ( \widehat{\phi}_1 F_{\mu\nu}^n + 
\sqrt{\widehat{\phi}_1 \widehat{\phi}_3} 
B_{\mu\nu}^n \right ) = 0 \quad , \eeeq
\beeq
  \label{4.6be}
2\sqrt{\widehat{\phi}_1 \widehat{\phi}_3} F_{\mu\nu}^n + 2 
\widehat{\phi}_3 B_{\mu\nu}^n - {h \over 2}
\varepsilon_{\mu\nu\rho\sigma} 
\partial_{\rho}\partial_{\lambda}\widetilde{B}_{\lambda\sigma}^n - 
{\sigma
\over 2} \left ( \partial_{\mu}\partial_{\sigma} B_{\sigma\nu}^n - 
\partial_{\nu} \partial_{\sigma}
B_{\sigma\mu}^n \right ) = 0 \ .
\emini

As dual fields we introduce vectors $C_{\mu}^n$ (whose field strength 
tensor will be denoted by
$F_{\mu\nu}^{c,n}$), and pseudoscalars $\varphi^n$. They are related 
to the original fields $A_{\mu}^n$,
$B_{\mu\nu}^n$ through

\bminiG{4.7e}
\label{4.7ae}
{1 \over 2}\ F_{\mu\nu}^{c,n} = \widehat{\phi}_1  
\widetilde{F}_{\mu\nu}^n + \sqrt{\widehat{\phi}_1 
\widehat{\phi}_3} \widetilde{B}_{\mu\nu}^n \quad , \eeeq
\beeq
  \label{4.7be}
\partial_{\mu} \varphi^n + C_{\mu}^n = {h \over 2} \ 
\sqrt{{\widehat{\phi}_1 \over \widehat{\phi}_3}} \  
\partial_{\mu} \widetilde{B}_{\mu\nu}^n
\emini

\noi (recall our definition of $\widetilde{B}_{\mu\nu}^n$ etc. in eq. 
(\ref{2.14e})). \par

The Bianchi identities associated to the dual fields are

\bminiG{4.8e}
\label{4.8ae}
\partial_{\mu} \widetilde{F}_{\mu\nu}^{c,n} = 0 \quad , \eeeq
\beeq
  \label{4.8be}
\varepsilon_{\mu\nu\rho\sigma}\ \partial_{\rho} \partial_{\sigma} 
\varphi^n = 0\quad .
\emini

\noi Contracting the dual of (\ref{4.7ae}) with $\partial_{\mu}$ one 
realizes immediately that the Bianchi
identity (\ref{4.8ae}) coincides with the equation of motion 
(\ref{4.6ae}). The Bianchi identity (\ref{4.8be})
should reproduce the equation of motion (\ref{4.6be}) up to the 
``gauge fixing term'' $\sim \sigma$. Indeed
this is the case up to terms proportional to $\partial_{\mu} 
\widehat{\phi}_i$. \par

The Bianchi identities associated with the original fields 
$A_{\mu}^n$, $B_{\mu\nu}^n$ read

\bminiG{4.9e}
\label{4.9ae}
\partial_{\mu} \widetilde{F}_{\mu\nu}^{n} = 0 \quad , \eeeq
\beeq
  \label{4.9be}
\partial_{\mu} \partial_{\nu} \widetilde{B}_{\mu,\nu}^n = 0\quad .
\emini

As action for the dual fields $C_{\mu}^n$, $\varphi^n$ we propose (up 
to gauge fixing terms)

\beq
\label{4.10e}
\Gamma_{Dual} = \int d^4x \left \{ {1 \over 4 \sqrt{\widehat{\phi}_1 
\widehat{\phi}_3}} F_{\mu\nu}^{c,n} \
F_{\mu\nu}^{c,n} + {2 \over h} \sqrt{{\widehat{\phi}_3 \over 
\widehat{\phi}_1}} \left ( \partial_{\mu}
\varphi^n + C_{\mu}^n \right )^2 \right \} \ . \eeq

\noi Now one finds, using (\ref{4.7be}), that the Bianchi identity 
(\ref{4.9be}) coincides exactly with the
equation of motion for $\varphi^n$ derived from (\ref{4.10e}), 
whereas the Bianchi identity (\ref{4.9ae})
agrees with the equation of motion of $C_{\mu}^n$ only up to terms 
$\sim \partial_{\mu} \widehat{\phi}_i$. \par

Not astonishingly, duality is thus realized to the same level as the 
gauge symmetry (\ref{4.2e}): To the order  quadratic in the fields the
physical content of the dual action  (\ref{4.10e}) is the same as the
original effective action (\ref{4.1e}), since the ``perturbing''  terms
$\sim \partial_{\mu} \widehat{\phi}_i$ are effectively of higher order.
(As solutions of the stationary point equations (\ref{3.52e}) the
deviations of the fields $\widehat{\phi}_i$ from their constant values
in the vacuum are proportional to the bilinear operators ${\cal
O}_i$.) To higher order in the fields duality is realized only on
modes of $A_{\mu}^n$, $B_{\mu\nu}^n$ which correspond to constant
bilinears ${\cal O}_i$. \par

The physical interpretation of the dual action (\ref{4.10e}) is 
obviously the one of an abelian $U(1)^N$
Higgs model in the spontaneously broken phase where $\varphi^n$ 
represent the Goldstone bosons, and where
the gauge fields $C_{\mu}^n$ have acquired a mass 
$2(\widehat{\phi}_3/h)^{1/2}$. Since this represents the
``low energy effective action'' of a theory in which the ``dual'' 
electric charge has condensed in the
vacuum, the original action (\ref{4.1e}) corresponds to the situation 
where the ``magnetic'' charge has
condensed in the vacuum. \par

Let us return to the response of our model with respect to external 
sources. The expression for $W(J)$ has
been given in eq. (\ref{3.42e}) in the preceeding section, and 
subsequently we consider the limit of
vanishing infrared cutoff $k^2 \to 0$. We recall that the fields 
$\widehat{\phi}_i(J)$ satisfy the
stationary point equations (\ref{3.44e}), and satisfy thus obviously

\beq
\label{4.11e}
\widehat{\phi}_i(J) = \widehat{\phi}_i^0 + {\cal O}(J^2)
\eeq

\noi (where the $\widehat{\phi}_i^0$ satisfy (\ref{3.49e})). \par

In fact, the procedure to solve (\ref{3.44e}) exactly would be quite 
cumbersome, since one would need the
expressions for the propagators $P^{rs}(x_1, x_2, \phi_i)$ for 
arbitrary $\phi_i$. In practice it is much
easier to work with the effective action $\Gamma (A,B)$ 
(\ref{3.51e}); this procedure will be
described below. \par

On the other hand, if the sources $J$ are sufficiently weak, we can 
expand $W(J)$ in powers of $J$. Since the
$\widehat{\phi}_i^0$ are stationary points of $G(\phi_i)$ we have

\beq
\label{4.12e}
W(J) = NG(\widehat{\phi}_i^0) - {1 \over 2} \int d^4x_1 \ d^4x_2 
\left \{ J_r^n(x_1)\ P^{rs}(x_1,x_2,
\widehat{\phi}_i^0) J_s^n(x_2) \right \} + {\cal O}(J^4) \ . \eeq

Interesting informations can already be obtained from the second term
of ${\cal O}(J^2)$ in eq. (\ref{4.12e}), which involves the propagators
$P^{rs}$ in a confining vacuum given in eqs. (\ref{A.7e}) and
(\ref{A.8e}) in the appendix. \par

Let us start with a source $J_{A,\mu}^n(x)$ for the fields  $A_{\mu}^n$
only. The simplest geometrical configuration is a Wilson loop source
$J_{A,\mu}^n$ which is non-vanishing only  on a curve $C$ embedded in
4d space-time, where $C$ has to be closed because of current
conservation:

\beq
\label{4.13e}
J_{A,\mu}^n(x) = ig_A \int_C dx'_{\mu} \ \delta^4(x - x') \quad .
\eeq

\noi The term quadratic in $J$ in (\ref{4.12e}) then becomes

\beq
\label{4.14e}
{Ng_A^2 \over 2} \int_C dx_{1,\mu} \int_C dx_{2,\nu} \ 
P_{\mu,\nu}^{AA}(x_1 - x_2)
\eeq

\noi with $P^{AA}$ as in eq. (\ref{A.7e}). The divergent constants 
in (\ref{A.7e}) disappear actually in the
expression (\ref{4.14e}). This can be seen most easily by applying 
the (abelian) Gauss law, and expressing
the source $J_{A,\mu}^n$ in terms of a source $J_{F,\mu\nu}^n$ which 
is non-vanishing on an (arbitrary) surface
$S$ bounded by the contour $C$:

\beq
\label{4.15e}
J_{A,\mu}^n(x) = 2 \partial_{\nu} J_{F,\nu \mu}^n(x)
\eeq

\noi with

\beq
\label{4.16e}
J_{F,\mu\nu}^n(x) = ig_F \int_S d^2\sigma_{\mu\nu}(z) \ \delta^4(x - z)
\eeq

\noi where $g_F = g_A/2$, $z$ parametrizes the surface $S$ and

\beq
\label{4.17e}
d^2\sigma_{\mu\nu}(z) = \varepsilon_{\mu\nu\rho\sigma} \ 
n_{\rho}^1(z) \ n_{\sigma}^2(z)\ d^2z \quad .
\eeq

\noi $n_{\rho}^1$, $n_{\sigma}^2$ are two orthogonal unit vectors
perpendicular  to $S$. \par

The expression (\ref{4.14e}) then turns into

\beq
\label{4.18e}
2N g_F^2 \int_S d^2\sigma_{\mu\nu}(z) \int_S d^2\sigma_{\rho\sigma}(z') \ 
P^{FF}_{\mu\nu, \rho\sigma} (z-z')
\eeq

\noi with $P^{FF}$ as in eq. (\ref{A.8ae}). The dependence on the 
divergent constants in eq. (\ref{A.7e}) has
disappeared as announced. \par

In the limit where the surface $S$ becomes very large one finds that
the expression  (\ref{4.18e}) is proportional to the (minimal) surface
bounded by the contour $C$ which we will also  denote by $S$. \par

Thus we have obtained the area law for the expectation value of the 
Wilson loop. This result has to be swallowed, however, with two grains
of salt. First, in the limit  where $S$ and hence the expression
(\ref{4.18e}) become large, it becomes inconsistent to confine  oneself
to terms of ${\cal O}(J^2)$ in $W(J)$. (Fortunately, as we will discuss
below, a more complete treatment just modifies the contributions to the
string tension, not the area law.) \par

Second, and more importantly, an inconsistency arises if we couple  our
model to a quantum field theory. In order
to clarify this inconsistency we first consider a slightly modified
configuration of sources $J_F$: Let us consider two distinct  ``Wilson
surfaces'' $S_1$, $S_2$, which may both be small and centered at $x_1$,
$x_2$, respectively. Accordingly the  source $J_F$ becomes the sum of
two terms, $J_F = J_{F,1} + J_{F,2}$, where $J_{F,1}$ and $J_{F,2}$
are  non-vanishing only near $x_1$ or $x_2$. Inserting $J_F$ into the
term quadratic in $J$ in $W(J)$ in  (\ref{4.12e}) one obtains, apart
from ``self-contractions'', a mixed term which is of the order of 
$P_{\mu\nu,\rho\sigma}^{FF}(x_1 - x_2)$ for $|x_1 - x_2|$ large
compared to the diameters of the Wilson surfaces. (The Lorentz indices
depend on the orientations of the surfaces $S_1$ and $S_2$.) Due to
the second term in (\ref{A.8ae}) $P^{FF}(x_1 - x_2)$ decreases only as 
$|x_1 - x_2|^{-2}$ for generic relative orientations. \par

The resulting expression can be interpreted as a contribution to the 
action of the configuration: The
effective action $\Gamma$ is given by the Legendre transform 
(\ref{3.50e}), and up to ${\cal O}(J^4)$ the
expression for $\Gamma$ thus coincides with the expression 
(\ref{4.12e}) for $W$ up to a change of the sign of
the second term. \par

At this point the result is not yet problematic. Let us now consider 
the coupling of our model to a quantum field theory in the form of
$\widetilde{J}_{F,\mu\nu}^n F_{\mu\nu}^n$, where the source term 
$\widetilde{J}_{F,\mu\nu}^n$ is quadratic in "external" quantum
fields.  ($SU(N_c)$ Yang-Mills theory, e.g., contains couplings of the
form $f_{nab}F_{\mu\nu}^n W_{\mu}^a W_{\nu}^b$, which corresponds to 
$\widetilde{J}_{F,\mu\nu}^n = f_{nab}W_{\mu}^a  W_{\nu}^b$.) Vacuum
bubbles of the quantum fields then correspond to a ``background'' of
sources $\widetilde{J}_{F,\mu\nu}^n$, which fills the whole space-time
(to be averaged over $\widetilde{J}_{F,\mu\nu}^n$ with, e.g., a
Gaussian measure). If the contribution to the action due to the induced
interactions between two vacuum bubbles decreases only as the
(distance)$^{-2}$ as described above, space-time filled with vacuum 
bubbles will lead to an infinite action. Even more severely, every
additional Wilson loop switched on  ``by hand'' will also lead to an
additional divergent contribution to the action due to its 
interactions with the vacuum bubbles (after averaging over
$\widetilde{J}_{F,\mu\nu}^n$ this divergence is only logarithmic at
large distances). Hence it seems to be disallowed\footnote{Note the
important difference to ordinary QED: Here the contribution to the
Euclidean action induced by two vacuum bubbles, i.e. two  dipoles
localized both in space and time, decreases as the (distance)$^{-4}$.
``Switching on'' an additional single dipole immersed in a bath of
disoriented dipoles then only costs a finite amount of action.}.\par

Actually, before we have assumed that we have obtained our model as 
an effective low energy theory {\it
after} the fields $W_{\mu}^a$ have been integrated out. The present 
paradox persists, however, if
previously we have only integrated over $W_{\mu}^a$-modes with $p^2 \ 
\gsim \ k_W^2$ (with $k_W^2$ as
small as one likes, but $\not= 0$), and the remaining modes with $p^2 
 \lsim \ k_W^2$ generate the
vacuum bubbles described above. In addition, in full QCD we have more 
fields like quarks, whose
vacuum bubbles will generate a background density of sources 
$\widetilde{J}_{F,\mu\nu}^n$. \par

The fact that we assumed that $W_{\mu}^a$-modes with $p^2 \ \gsim \ 
k_W^2$ have already been
integrated out will, on the other hand, lead to a solution of the 
problem: First, in the original
$SU(N_c)$ Yang-Mills theory, a gauge-invariant Wilson loop ``source'' 
is coupled both to $A_{\mu}^n$
and to $W_{\mu}^a$. If we integrate over $W_{\mu}^a$-modes after 
having introduced the fields
$B_{\mu\nu}^n$ as in eqs. (\ref{2.5e}) and (\ref{2.6e}), this will 
necessarily generate a coupling
of the Wilson loop source to $B_{\mu\nu}^n$. Hence it is inconsistent 
to switch on a source for
$F_{\mu\nu}^n$, but not simultaneously a source for $B_{\mu \nu}^n$. 
The simplest expression for a
source for $B_{\mu\nu}^n$ is again of the form (\ref{4.16e})

\beq
\label{4.19e}
J_{B,\mu\nu}^n(x) = ig_B \int_S d^2\sigma_{\mu\nu}(z) \ \delta^4(x - z) \quad ,
\eeq

\noi where the surface $S$ coincides with the one in (\ref{4.16e}). 
The value of the coupling constant $g_B$
can, a priori, not be predicted within our model (neither the 
couplings $g_F$ or $g_A$ in (\ref{4.16e}) or
(\ref{4.13e})). \par

Thus we return to the term quadratic in $J$ in (\ref{4.12e}), and 
insert for the source $J$ a sum $J_F + J_B$
with $J_F$ as in (\ref{4.16e}), and $J_B$ as in (\ref{4.19e}). With 
the propagators from (\ref{A.7e}) and
(\ref{A.8e}) this term becomes

\bea \label{4.20e}
&&{1 \over 16 \pi^2\sigma} \int_S d^2\sigma_{\mu\nu}(z_1) \int_S 
d^2\sigma_{\rho\sigma}(z_2) \Big \{ \Big (
\sqrt{{\phi_3 \over \phi_1}} g_F - g_B \Big )^2 T_{1, 
\mu\nu,\rho\sigma}(\partial_{z_1})\log |z_1 -
z_2|\nn \\
&&  + {\cal O} \left ( |z_1 - z_2|^{-4} \right ) \Big \} \ .\eea

\noi Thus the terms of ${\cal O}(|z_1 - z_2|^{-2})$ in the curled 
parenthesis in (\ref{4.20e}) vanish if

\beq
\label{4.21e}
\sqrt{\phi_3} \ g_F = \sqrt{\phi_1} \ g_B \quad .
\eeq

Again, at this point this is not an obligatory constraint. However, 
if we switch on a) ``Wilson loop'' sources
$J_F$, $J_B$, b) simultaneously uncorrelated ``background sources'' 
$\widetilde{J}_F$, $\widetilde{J}_B$, c)
average over $\widetilde{J}_F$, $\widetilde{J}_B$ with an arbitrary 
Gaussian measure, we obtain a
logarithmically infrared divergent expression, unless the condition 
(\ref{4.21e}) on $J_F$, $J_B$ is
satisfied. Hence, in the background of a fluctuating vacuum it costs 
infinite action to turn on sources $J_F$,
$J_B$ which are not related through (\ref{4.21e}). This constraint 
can be interpreted as a constraint on the
ratio $\phi_1/\phi_3$ in the presence of sources, or on the ratio 
$g_F/g_B$ (once $g_F$ and $g_B$  vary, e.g.,
with the vacuum configuration of a ``microscopic'' theory) or, most 
realistically, on a combination of
both.\par

Before discussing the resulting behaviour of the expectation value of 
the Wilson loop we note that the above
consideration leads, in general, to the following constraint on 
sources $J_A$, $J_B$ or $J_F$, $J_B$:

\bminiG{4.22e}
\label{4.22ae}
\sqrt{\phi_3} \ J_{A,\mu}^n(x) = 2 \sqrt{\phi_1}\ \partial_{\nu}  
J_{B,\nu\mu}^n(x) \quad \hbox{or}\eeeq
\beeq
  \label{4.22be}
\sqrt{\phi_3} \ J_{F,\mu\nu}^n(x) =  \sqrt{\phi_1} \ J_{B,\mu\nu}^n(x)\quad .
\emini

\noi The constraints (\ref{4.22e}) are obviously related to 
the gauge symmetry (\ref{4.2e}): If we
replace $\phi_i$ by the vacuum configurations (in the absence of 
sources) $\widehat{\phi}_i$, (\ref{4.22e})
corresponds to the ``current conservation condition'', i.e. to the 
condition that

\beq
\label{4.23e}
\int d^4x \left ( J_{A,\mu}^n \ A_{\mu}^n + J_{B,\mu\nu}^n \ 
B_{\mu\nu}^n \right )
\eeq

\noi is invariant under (\ref{4.2e}). \par

In the case of ``conventional'' gauge symmetries the current 
conservation conditions can (and have to) be
imposed by hand in order to ensure renormalizability and unitarity of 
the theory. In the present model, on the
one hand, current conservation cannot be imposed from the beginning, 
since the associated (approximate) gauge
symmetry appears only at the level of the effective action once the 
equations of motion of the fields $\phi_i$
are satisfied. On the other hand, renormalizability is not an issue 
here, since we consider an effective low
energy theory with a fixed UV cutoff. Also unitarity is trivial as 
long as we assume the absence of poles in
the propagators in the Minkowskian regime, cf. the discussion before 
eq. (\ref{2.15e}). Thus the model has
no $S$-matrix at all, i.e. no asymptotic states. (The absence of 
bound states will briefly be discussed
below). Consequently at this level no constraints on the sources arise. \par

Only if we couple the model to a quantum field theory we have to 
reconsider the question of unitarity, i.e.
the possibility to project -- in a Lorentz covariant way -- onto a 
positive semi-definite part of its
Hilbert space which is represented, loosely speaking, by our sources 
$J$ viewn as functionals of fields.
Precisely in this situation ``vacuum bubbles'' impose eqs. 
(\ref{4.22e}), the analog of current
conservation in an abelian Higgs model with an additional 
Lorentz index attached to the currents.
\par

If eqs. (\ref{4.22e}) hold the expectation value of the Wilson loop 
has to be reconsidered: The Wilson loop
corresponds now to a source $J_A$ of the form (\ref{4.13e}), which 
can be rewritten as a source $J_F$ of the
form (\ref{4.16e}) using (\ref{4.15e}), {\it plus} a source $J_B$ of 
the form (\ref{4.19e}) where the
surface $S$ coincides with the one in (\ref{4.16e}). \par

First, the term quadratic in $J$ in (\ref{4.12e}) consists now only 
of the ``short range'' contributions
neglected in (\ref{4.20e}). Using again the propagators (\ref{A.7e}) 
and (\ref{A.8e}) these read

\bea
\label{4.24e}
&&\int_S d^2\sigma_{\mu\nu}(z_1) \int_S d^2\sigma_{\rho\sigma}(z_2) 
{-g_F^2 \over 16 \pi^2 \phi_1} \sqrt{{\phi_3
\over h}}  \nn \\
&& \times \left ( T_{1,\mu\nu,\rho\sigma}(\partial )
- {4 \phi_3 \over h} T_{2,\mu\nu ,\rho\sigma} \right ) 
{1 \over |z_1 - z_2|} K_1\left ( 2 |z_1 - z_2| \sqrt{{\phi_3 \over 
h}} \right ) \ .  \eea

\noi With (\ref{4.21e}) the contributions $\sim |z_1 - z_2|^{-4}$ 
cancel as well, and we recall that the
tensors $T_1$ and $T_2$ are given in (\ref{A.2e}) in the appendix.  \par

In the limit where the surface $S$ becomes large the expression 
(\ref{4.24e}) behaves as
 
\beq
\label{4.25e}
S \cdot {2g_F^2 \over \pi \phi_1} \left ( {\phi_3 \over h}\right 
)^{3/2} \int_0^{\infty} dz \ K_1\left (
2z\sqrt{{\phi_3 \over h}} \right )\quad . \eeq

\noi Hence it implies the area law in spite of the cancellations of
the  long range contributions of the propagators. (Since we have
omitted the UV cutoffs in the space-time  propagators the expression
(\ref{4.25e}) is seemingly UV divergent.) \par

Expressions of the form (\ref{4.25e}) for the (negative) exponent of 
the expectation value of the Wilson loop
have already appeared repeatedly in the literature in the context of 
the method of vacuum correlators
\cite{3r,20r}: In the Gaussian approximation (the stochastic vacuum 
model) the expectation value of the
Wilson loop is given by the expectation value of the field strength 
correlator, and many models for this
correlator lead to (\ref{4.25e}) \cite{6r,14r}. Actually already the 
previous result (\ref{4.18e}) -- the
area law for Wilson loop sources $J_A$ only -- has an interpretation 
in this approach: It would correspond
to a function $D_1(x^2)$ in the standard decomposition of the field 
strength correlator \cite{3r,6r,20r}
which decreases only as $|x|^{-2}$ for large $|x|$; such a behaviour 
also implies the area law, but it is
strongly disfavoured both phenomenologically \cite{5r} and from 
lattice data \cite{21r}. \par

As we have already stated several times above, sources for operators 
in a $SU(N_c)$ Yang-Mills theory correspond in our model -- if
considered as an ``effective low energy  theory'' -- to a priori
unknown superpositions of sources for $A_{\mu}^n$ (or $F_{\mu\nu}^n$)
and  $B_{\mu\nu}^n$. The previous discussion leading to
eqs.~(\ref{4.22e}) fixes this ambiguity, and precisely in this case
the two-point correlators decrease exponentially in agreement with the
$SU(N_c)$ lattice data. \par

In our model the expression (\ref{4.25e}) for the resulting  string
tension can, however, not be taken too seriously, since in the limit of
a large surface $S$ the terms  of higher order in $J$ in $W(J)$ in
(\ref{4.12e}) can no longer be neglected. \par

In order to compute $W(J)$ in the ``non-linear'' regime one has 
to start first with the effective action
$\Gamma (A, B)$ as given in eq. (\ref{3.51e}) (in the limit of 
vanishing infrared cutoff $k$) or eq.
(\ref{4.1e}). One has to solve the combined equations of motion for 
$A_{\mu}^n$, $B_{\mu\nu}^n$ and $\phi_i$
in the presence of sources $J_{A,\mu}^n$ and $J_{B,\mu\nu}^n$: The 
corresponding equations for $A_{\mu}^n$ and
$B_{\mu\nu}^n$ are given by eqs. (\ref{4.6e}) (where, however, the 
zeros on the right-hand sides have to be
replaced by the corresponding source terms and 
$2 \sqrt{\widehat{\phi}_1 \widehat{\phi}_3}$ has to be replaced
by $\widehat{\phi}_2$). The equations for $\phi_i$ are given by eqs. 
(\ref{3.52e}) or (\ref{3.54e}), where
the operators ${\cal O}_i$ depend on $F_{\mu\nu}^n$ and 
$B_{\mu\nu}^n$ as in (\ref{2.12e}). \par

The corresponding fields have to be inserted into $\Gamma (A,B)$ in 
(\ref{4.1e}), and then one has to ``undo''
the Legendre transformation (\ref{3.48e}) in order to obtain $W(J)$. 
As already noted this last step is
actually trivial since $\Gamma (A,B)$ is quadratic in $A$, $B$: It 
suffices to change the sign of the second
term in the expression (\ref{4.1e}). Then one can study the 
dependence of $W(J)$ on $J$ in its full beauty. \par

Albeit generally the solution of the combined equations of motion is 
certainly quite involved, there is one situation where it becomes
straightforward: Let us assume that we  have turned on sources
$J_{F,\mu\nu}^n(x)$ and $J_{B,\mu\nu}^n(x)$ which are related as in eq.
(\ref{4.22be}),  and which are constant inside a space-time volume $V$,
and vanishing outside. Let us furthermore assume that we  can neglect
all derivative terms in $G(\phi_i)$. Then we can allow for constant
configurations  $\widehat{\phi}_i^{V}$ inside the volume $V$, and
constant (generally different) configurations $\widehat{\phi}_i^0$ 
outside $V$; the discontinuities at the boundaries of $V$ cost no
energy in this case. The arguments leading  to eqs. (\ref{4.22e}) are
based on the long distance behaviour of the propagators in the vacuum
outside $V$,  accordingly eqs. (\ref{4.22e}) have to hold for
$\widehat{\phi}_i^0$ outside $V$. For simplicity  we assume that they
also hold for $\widehat{\phi}_i^{V}$ inside $V$, i.e. that
$\widehat{\phi}_1^{V}/\widehat{\phi}_3^{V} =
\widehat{\phi}_1^0/\widehat{\phi}_3^0$.  \par

For the fields $F_{\mu\nu}^n(x)$ and $B_{\mu\nu}^n(x)$ the following 
simple solution is then possible: For
$B_{\mu\nu}^n(x)$ one chooses

\beq
\label{4.26e}
B_{\mu\nu}^n(x) = 0 \eeq

\noi everywhere, both inside and outside $V$ (because of the 
derivatives acting on $B_{\mu\nu}^n$ in the
action discontinuities in $B$ are not allowed). Then eqs. 
(\ref{4.6e}) -- with source terms included and
using the ``confining'' relation $\phi_2 = 2\sqrt{\phi_1\phi_3}$ both 
inside $V$ and outside $V$ -- collapse
to

\bminiG{4.27e}
\label{4.27ae}
2 \partial_{\mu} \Big (\phi_1 F_{\nu\mu}^n(x) \Big ) = \partial_{\mu} 
J_{F,\mu\nu}^n(x) \quad , \eeeq
\beeq
\label{4.27be}
2\sqrt{\phi_1\phi_3} F_{\mu\nu}^n(x) = J_{B,\nu\mu}^n(x)\quad .
\emini

\noi With $J_F$ and $J_B$ are related as in (\ref{4.22be}) inside $V$ 
eqs. (\ref{4.27e}) are trivially solved
simultaneously by

\bminiG{4.28e}
\label{4.28ae}
F_{\mu\nu}^n(x) = {1 \over \widehat{\phi}_1^{V}} \ J_{F,\mu\nu}^n(x) 
\quad \hbox{inside \ V} \quad ,\eeeq
\beeq
\label{4.28be}
F_{\mu\nu}^n(x) = 0 \qquad \hbox{outside\ V\ (where}\ J_F = J_B =
0)\quad .
\emini

Hence, in this simple scenario the effective action (\ref{4.1e}) is
given by
 
\beq
\label{4.29e}
\Gamma = \int_V d^4x \left \{ N \widetilde{G}(\widehat{\phi}_i^V) + 
\widehat{\phi}_1^V {\cal O}_1 \right \}
+ \int_{\bar{V}} d^4x \ N \widetilde{G}(\widehat{\phi}_i^0) \eeq

\noi where $\widetilde{G}(\phi_i)$ is the density associated to 
$G(\phi_i)$, and $\bar{V}$ denotes  the entire
space-time volume outside $V$. (Recall that only ${\cal O}_1$ is
non-zero for $B_{\mu\nu}^n = 0$.) Since the configurations 
$\widehat{\phi}_i^0$ are local maxima of
$\widetilde{G}(\phi_i)$ and the configurations 
$\widehat{\phi}_i^V$ are local maxima of
$\widetilde{G}(\phi_i) + \phi_1{\cal O}_1$ one derives easily

\beq
\label{4.30e}
N \widetilde{G}(\widehat{\phi}_i^V) + \widehat{\phi}_1^V {\cal 
O}_1 > N \widetilde{G}(\widehat{\phi}_i^0)
\quad . \eeq

Hence the presence of constant sources inside a volume $V$ increases 
the Euclidean action by an amount which
is proportional to $V$ (in an approximation where derivative terms in 
$G(\phi_i)$ are neglected). \par

This consideration can be applied to the Wilson loop -- i.e. sources 
$J_F$, $J_B$ given by (\ref{4.16e}) and
(\ref{4.19e}) with $g_F$ and $g_B$ related by 
$\sqrt{\widehat{\phi}_3^{V}} g_F =
\sqrt{\widehat{\phi}_1^{V}}g_B$ -- provided we replace the enclosed 
surface $S$ by a ``layer'' of finite
thickness: In the presence of an UV cutoff $\Lambda$ we cannot 
resolve distances smaller than $\Lambda^{-1}$
anyhow, and hence we should replace the surface $S$ in (\ref{4.16e}) 
and (\ref{4.19e}) by a layer of
thickness $\Lambda^{-1}$ and a corresponding volume $V = S \cdot 
\Lambda^{-1}$. \par

Given the above discussion one then obtains immediately an action 
$\Gamma$ which is proportional to $S$, and
-- after a Legendre transform -- $W(J)$ proportional to $S$ and thus 
the area law. In the present scenario the
profile of the flux tube associated to the Wilson loop would be 
discontinuous, with $\phi_i =
\widehat{\phi}_i^V$ (and $F_{\mu\nu}^n$ given by (\ref{4.28ae})) 
inside the flux tube, and $\phi_i =
\widehat{\phi}_i^0$ outside. If we would take derivative terms in 
$G(\phi_i)$ into account, this profile would
be smoothened, and additional contributions to the action 
proportional to the length of the loop $C$ would be
obtained. As stated above, a treatment of these effects would require 
the solution of the combined equations
of motion of $\phi_i$, $F_{\mu\nu}^n$ and $B_{\mu\nu}^n$ (in some 
similarity to the much simpler scenario
\cite{6r} based on the dual abelian Higgs model) and depend evidently 
on the unknown bare action $S_{bare}$
in (\ref{3.33e}) or $G_{bare}$ in (\ref{3.35e}), (\ref{3.37e}). \par

In order to close this section we briefly report on our fruitless 
search for bound states in the model in the
confining phase. In principle these would show up as poles in the 
propagators of the fields $\phi_i$ for
$q^2 < 0$. To this end one has to develop the fields $\phi_i$ around 
the vacuum configuration,

  \beq
\label{4.31e}
\phi_i(x) = \widehat{\phi}_i^0 + \delta \phi_i(x) \quad .
\eeq

\noi Inserting (\ref{4.31e}) into the action (\ref{4.1e}) the 
propagators for $\phi_i$ are obtained from the
terms quadratic in $\delta \phi_i$. It is sensible to study first the 
``mass matrix'' $M_{ij}^2$ in the space
$\delta \phi_i$:

\beq
\label{4.32e}
M_{ij}^2 = {N \over 2} \ {\delta^2 G(\phi_i) \over \delta \phi_i \ 
\delta \phi_j} \left |_{\widehat{\phi}_i^0}
\right . \quad .  \eeq

\noi We recall that $G(\phi_i)$ is composed out of $G_{bare}$ and 
$\Delta G$, cf. eq. (\ref{3.43e}), with
$\Delta G$ given in (\ref{3.41e}) for $k^2 \to 0$, and the 
$\widehat{\phi}_i^0$ satisfy (\ref{3.49e}). Due to
the non-analytic behaviour of $\Delta G$ in the confining phase one 
finds that all entries of $M_{ij}^2$
actually diverge for $k^2 \to 0$, but $M_{ij}^2$ has two finite 
eigenvalues (depending on $G_{bare}$) with
eigenvectors given by

\beq
\label{4.33e}
\left ( \begin{array}{l} \sqrt{\widehat{\phi}_1^0} \\ 
\sqrt{\widehat{\phi}_3^0} \\ \quad 0\end{array} \right )
\delta \phi_a \quad , \quad \left ( \begin{array}{l} \quad 0 \\ 
\sqrt{\widehat{\phi}_1^0} \\
\sqrt{\widehat{\phi}_3^0}  \end{array} \right ) \delta \phi_b \quad .
\eeq

\noi We have computed the full momentum dependence of the contribution 
$(\Delta P^{-1}(q^2))_{ab}$ to the inverse
propagators, which arises from the quantum contribution $\Delta G$ 
(\ref{3.39e}) after expanding $\phi_i$ to
second order in $\delta \phi_i$ in the ``directions'' specified by 
(\ref{4.33e}), for $|q^2| \ \lsim \
\Lambda^2$. We found that $(\Delta P^{-1}(q^2))_{ab}$ depend only 
weakly (logarithmically) on $q^2$, and
that its determinant is negative definite. Together with the 
negativity of $G_{bare}(\phi_i)$ (recall the
remark below eq. (\ref{3.35e})) this implies that the full inverse 
propagators do not vanish for $|q^2|\ \lsim \ \Lambda^2$, thus no
bound-state pole appears at least below the UV cutoff. \par

Such a pole would actually be a disaster for unitarity: The 
negativity of $G(\phi_i)$, notably for
$|\phi_i| \to \infty$, is harmless only if the fields $\phi_i$ are 
pure ``auxiliary fields'', i.e. with
algebraic equations of motion. \par

On the other hand one may be deceived by the absence of 
``glueballs'', if the present model is interpreted as
an effective low energy theory for $SU(N_c)$ Yang-Mills theory. 
However, we recall that the present model
would only describe the abelian subsector of $SU(N_c)$ Yang-Mills in 
the MAG, and that we finally have to add
the (massive) ``non-abelian'' gluons $W_{\mu}^a$ as well. Our theory 
induces flux-tube like confining
attractive interactions among all fields which couple like 
$J_{F,\mu\nu,}^nJ_{B,\mu\nu}^n$ to $F$ and $B$,
hence the $W_{\mu}^a$-gluons will necessarily form bound states which 
will correspond to the desired
glueballs. (This scenario is quite consistent with heavy glueballs, 
given $M_W \sim {\cal O}$(1 GeV) in the
MAG \cite{12r}). \par

Further details on the possible relation between the present model 
and $SU(N_c)$ Yang Mills will be
discussed in the next section.

\mysection{Discussion and Outlook}
\hspace*{\parindent}  In this final section we want to discuss some 
general features of the model, notably the
relation between the large $N$ limit and $SU(N_c)$ Yang Mills at 
large $N_c$, and the systematic inclusion of
higher derivatives in the ``bare''action. \par

First, we had emphasized the solvability of the $A_{\mu} - 
B_{\mu\nu}$ -- model for $N \to \infty$ in the sense
that $W(J)$ or $\Gamma (A,B)$ can be given explicitely for arbitrary 
$G_{bare}(\phi_i)$ as in (\ref{3.42e}) or
(\ref{4.1e}), and one is left with the coupled equations of motion 
for $A_{\mu}^n$, $B_{\mu\nu}^n$ and
$\phi_i$. To this end we had to assume a certain $N$-dependence of 
the coefficients of the bare action
$S_{bare}(F,B)$, cf. eq. (\ref{2.16e}):

\beq
\label{5.1e}
S_{bare}(F,B) = N \widehat{S}_{bare}\left ( {F \over \sqrt{N}} , {B 
\over \sqrt{N}}\right )
\eeq

\noi where the coefficients of $\widehat{S}_{bare}$ are
$N$-independent. \par

Let us now assume that $S_{bare}(F,B)$ has been obtained from 
$SU(N_c)$ Yang-Mills after integrating out the
``non-abelian'' gluons $W_{\mu}^a$ as in eq. (\ref{2.9e}). From a 
simple analysis of the contributing Feynman
diagrams and employing $\alpha_s \sim N_c^{-1}$ one obtains, in the
large $N_c$ limit, 

\beq
\label{5.2e}
S_{bare}^{YM}(F,B) = N_c^2 \ \widehat{S}_{bare}^{YM} \left ( {F \over 
\sqrt{N_c}}, {B \over \sqrt{N_c}} \right
) \eeq

\noi in contrast to (\ref{5.1e}). In principle it is still possible 
to introduce the operators ${\cal O}_i$
(\ref{2.12e}), and to express $exp(- S_{bare}^{YM}({\cal 
O}_i))$ in terms of a path integral over
$\phi_i$ as in (\ref{3.33e}):

\beq
\label{5.3e}
e^{-S_{bare}^{YM}({\cal O}_i)} = {1 \over {\cal N}} \int {\cal D} 
\phi_i \ e^{-N_c^2 G_{bare}(\phi_i/N_c) -
\int d^4x\phi_i{\cal O}_i} \ .\eeq

\noi The powers of $N_c$ on the right-hand side of (\ref{5.3e}) have 
been introduced such that, in the
stationary point approximation, the coefficients of $G_{bare}$ are of 
${\cal O}(1)$. (This becomes clear after
a rescaling $\phi_i \to N_c \phi_i$, ${\cal O}_i \to N_c {\cal O}_i$ 
and using (\ref{5.2e})). However, now the
stationary point approximation of the path integral (\ref{5.3e}) can 
no longer be justified in the large $N_c$
limit. \par

In addition, the structure of the operators ${\cal O}_i$  (\ref{2.12e})
assumes a global $O(N)$ symmetry which corresponds to a rotation in the
space of the $N = N_c - 1$ $U(1)$  subgroups of $SU(N_c)$ (or among the
fields $A_{\mu}^n$, $B_{\mu\nu}^n$, $n = 1 \dots N$). This $O(N)$
symmetry  is {\it not} a subgroup of $SU(N_c)$ and is generally not a
symmetry of $S_{bare}^{YM}(F,B)$. (It appears as an accidental global
symmetry, however, if one assumes that $S_{bare}^{YM}(F,B)$ contains
just quadratic $SU(N_c)$ invariants.) In general it is thus necessary
to introduce $N$ copies of each operator ${\cal O}_i$ in (\ref{2.12e}),

\beq
\label{5.4e}
{\cal O}_{i=1}^{(n)} = F_{\mu\nu}^n F_{\mu\nu}^n \quad \hbox{etc. \ 
for $i= 2, 3$}
\eeq

\noi {\it without} sum over $n$ on the right-hand side (or $N$ 
independent linear combinations of ${\cal O}_i^{(n)}$) \footnote{Note
that a $Z_2$ symmetry (charge conjugation  of all $W_{\mu}^a$ together
with $A_{\mu}^n \to - A_{\mu}^n$) {\it is} a subgroup of $SU(N_c)$,
which  allows to restrict ourselves to bilinear operators.} . In
particular for small $N$ the introduction of  operators ${\cal
O}_i^{(n)}$ is straightforward, but again the stationary point
approximation of the  path integral operators ${\cal O}_i^{(n)}$
(\ref{5.3e}) can no longer be supported by a large $N_c$ limit. \par

We emphasize, however, that the essential phenomena described by the 
present class of models do not depend on
the large $N$ limit (as the phenomenon of spontaneous symmetry 
breaking in bosonic $O(N)$ models): These are
the appearance of a confining phase, associated with an approximate 
(low energy) vector gauge symmetry and
duality transformation, and associated with the area law for the 
Wilson loop. They have their origin in the
convexity of a quantum effective action in spite of a non-convex bare 
action, which holds independently of a
large $N$ limit. Just the precise relation between the effective 
action and the bare action can only be
obtained for $N \to \infty$. An application of the present approach 
to $SU(N_c)$ Yang Mills will thus require
-- apart from a determination of $S_{bare}^{YM}$ as in eq. 
(\ref{2.9e}) -- a performance of the ${\cal
D}\phi_i$ path integrals beyond the stationary point approximation. \par

In general we cannot expect that a ``bare'' action as obtained by eq. 
(\ref{2.9e}) depends only on operators
${\cal O}_i$ of the form (\ref{2.12e}), since it will involve the 
most general terms with higher derivatives.
On the one hand we recall that the functional $G_{bare}(\phi_i)$ 
introduced in (\ref{3.34e}), which
specifies the models solved in Section 3, may well contain arbitrary 
derivatives acting on $\phi_i$. These
functionals $G_{bare}(\phi_i)$ correspond to bare actions 
$S_{bare}({\cal O}_i)$ with arbitrary derivatives
acting on ${\cal O}_i$. On the other hand, these do not correspond to 
the most general form of higher
derivative terms: In general it will be necessary to consider 
bilinear operators of the form

\beq
\label{5.5e}
{\cal O}_{i,\lambda_1\cdots \lambda_i} = F_{\mu\nu} \ 
f_{\mu\nu\rho\sigma , \lambda_1 \cdots \lambda_i} \
F_{\rho\sigma} \eeq

\noi (and with $F$ replaced by $B$) where $f$ depends on Kronecker 
symbols or $\varepsilon$-symbols in the
Lorentz indices {\it and} derivatives $\displaystyle{\mathrel{\mathop 
{\partial_{\mu}}^{\leftrightarrow}}}$
(with $F_1 \displaystyle{\mathrel{\mathop 
{\partial_{\mu}}^{\leftrightarrow}}} F_2 \equiv F_1
\partial_{\mu}F_2 - (\partial_{\mu} F_1)F_2)$. \par

Nevertheless it is possible to proceed as before: One can introduce 
auxiliary fields $\phi_{i,\lambda_1
\cdots \lambda_i}$ for each of the operators in (\ref{5.5e}), and 
rewrite the most general bare action
$S_{bare}({\cal O}_i)$ as

\beq
\label{5.6e}
e^{-S_{bare}({\cal O}_i)} = {1 \over {\cal N}} \int {\cal D} \phi_i \ 
e^{-G_{bare}(\phi_i) - \int
d^4x\phi_i{\cal O}_i} \eeq

\noi (adding, if one wishes, appropriate factors of $N$ as in 
(\ref{3.34e})). Of course the appropriate
contraction of Lorentz indices is understood in $\phi_i{\cal O}_i$. \par

Inserting (\ref{5.6e}) into (\ref{3.33e}) one obtains an expression 
for $W_k(J)$ analogous to (\ref{3.36e}),
and the ${\cal D}A{\cal D}B$ path integrals are still Gaussian. The 
essential technical complication is now
the construction of the propagators $P^{rs}$ of the $A_{\mu}$, 
$B_{\mu\nu}$ -- system which appear in
(\ref{3.38e}), (\ref{3.39e}), and the computation of $\Delta 
G(\phi_i)$ from (\ref{3.39e}) via a
generalization of (\ref{3.40e}) with higher powers of $p^2$ in the 
arguments of the logarithms. \par

However, one quickly realizes that the appearance of a confining 
phase, i.e. a solution of the stationary
point equations for $\phi_i$ of the form of eqs. (\ref{3.48e}) and 
(\ref{3.49e}) depends only on the
infrared behaviour of the propagators $P^{rs}(p^2)$ for $p^2 \to 0$. 
This, in turn, is unchanged by the
inclusion of operators ${\cal O}_i$ in (\ref{5.6e}) involving higher 
derivatives. \par

The general features of the model will thus remain untouched by the 
inclusion of more general operators of
the form (\ref{5.5e}), but its ``fine-structure'' will be affected: 
Lorentz scalar auxiliary fields
$\phi_i$ associated to Lorentz scalar operators ${\cal O}_i$ 
involving higher derivatives will generally
have non-vanishing vevs and affect the vacuum structure, and 
auxiliary fields $\phi_i$ with Lorentz indices
will affect the ``response'' of the model to sources $J_{F,\mu\nu}$, 
$J_{B,\mu\nu}$. \par

Since the inclusion of a large number of additional auxiliary fields 
$\phi_i$ is evidently a rather ambitious program\footnote{The
introduction of pseudoscalar operators of the  form
$\varepsilon_{\mu\nu\rho\sigma} F_{\mu\nu}F_{\rho\sigma}$ etc., without
higher derivatives, should  however be feasable. We would not expect
vevs of the associated auxiliary fields, which would spontaneously 
break parity, but their two point functions would give us informations
on the topological susceptibility.} the  question arises whether -- and
to what extent -- one could make use of the property of
``universality'' of  local quantum field theories as the ones presented
here which, by definition, corresponds to features which  are
practically independent of the UV cutoff $\Lambda$ and the bare action.
The phenomenon of universality  arises in the response of a quantum
field theory to external sources which induce only ``low energy'' 
phenomena: The momenta $q_i$ of the sources have to satisfy $q_i^2 \ll
\Lambda^2$, and the change $\Delta  E$ of the vacuum energy (in the
case of, e.g., constant sources) has to be much smaller than
$\Lambda$.  (Otherwise one is sensitive to the non-universal regime of
the effective potential.) \par

First, the Wilson loop corresponds to a source which satisfies none of
these criteria: Even after rewriting it as a source located on a
surface  $S$ bounded by the loop (as in (\ref{4.16e})) this source
varies rapidly in $x_{\bot}$ perpendicular  to $S$, and its expectation
value depends thus on the non-universal terms involving high
derivatives in  the effective action. Fortunately this does not affect
the area law but, as we have seen in section 4, the  computation of the
string tension. In addition, even after replacing an infinitely thin
surface $S$ by a  layer of finite thickness, the field configurations
inside this layer will generically depend on the  non-universal regime
of the effective action unless the ``source'' is unconventionally weak
(or ``diluted''). \par

The response of the present class of models to sources which vary
sufficiently slowly in $x$ and which are sufficiently weak will, on
the other hand, not depend on details of $G_{bare}$ and/or higher
derivative terms. Light quarks in QCD (with extended wave functions)
could play  the role of such sources, and their coupling to the present
model will constitute an interesting task in  the future\footnote{At
least a computation of $\Gamma_{eff}(\phi_i, A, B)$ in a systematic
expansion in  ($\partial_{\mu}/M_W$), beyond the large $N$ limit,
should be feasable.}. \par

To conclude, we have investigated a class of four-dimensional gauge 
theories with finite UV cutoff, which
exhibit confinement  and allow nevertheless -- in the large $N$ limit 
-- for controllable computations
notably in the infrared regime. Some features of the confining phase 
correspond quite to our expectations,
notably the possiblity to perform a duality transformation of the low 
energy part of the effective action
and thus to interpret confinement as monopole condensation. A 
technically related but nevertheless
unexpected feature is the appearance of a low energy vector gauge 
symmetry, which allows to ``gauge
away'' the low momentum modes of the abelian gluons $A_{\mu}^n$ and 
explains the absence of the
corresponding asymptotic states. If we assume in addition, as 
in eq. (\ref{2.15e}), higher
derivative terms bilinear in $B_{\mu\nu}^n$ such that the 
corresponding massive poles in the
propagators are absent, the model has no asymptotic states at all 
(even no bound states, cf. the end of
section 5). \par

The only ``meaning'' of the model is thus its reaction to external 
sources or currents. In the case of
conserved currents (with respect to the approximate vector gauge 
symmetry) confinement is then actually not
based on a long-range attractive force between two distinct sources 
-- the corresponding components of the
propagators cancel precisely for conserved external currents as 
explained in the paragraphs below eq.
(\ref{4.20e}) -- but on a different phenomenon: The minimal 
geometric object which can be associated to a
conserved current or source $J_{\mu\nu}$ is a surface $S$ (cf. the 
expressions (\ref{4.16e}) and
(\ref{4.19e})) and, as discussed in section 4, the contribution to 
the action of a source which is constant on
$S$ is proportional to $S$ (at least for large $S$) inducing the area 
law of the Wilson loop. The tensor
structure $T_{2,\mu\nu,\rho\sigma}$ in the propagator $P^{BB}$ (cf. 
eqs. (\ref{A.2e}), (\ref{A.7e})) is
essential for this result. The question why a pointlike particle like 
a heavy quark can be associated to
source for $B_{\mu\nu}$ which is constant on a surface $S$ is 
returned to the underlying microscopic
theory, such as the non-abelian sector of $SU(N_c)$ Yang-Mills, but here
we found that this is obligatory in order to be able to satisfy the
constraints (\ref{4.22e}) \par

Interestingly, if the lattice results \cite{10r} on the masses of the 
non-abelian gluons in the MAG are
correct, an elaboration of the precise relation between the present 
``effective low energy theory'' and
$SU(N_c)$ Yang Mills -- i.e. the structure of $S_{bare}^{YM}$ as well 
as the structure of the sources --
would involve only physics at comparatively small scales of ${\cal 
O}$((1 GeV)$^{-1}$) and is thus perhaps
feasable. \par

Apart from this evident application of our class of models we would 
like to point out that generalizations
in the sense of \cite{8r,22r} suggest itself: It is fairly 
straightforward to vary (notably increase) the
rank of the tensors $A_{\mu}$, $B_{\mu\nu}$ (maintaining rank (B) = 
rank (A) + 1) and to vary the dimension
of the space-time. This way one can develop dynamical models for the 
condensation of topological defects of
various dimensions ($p$ -- Branes) in various space-time dimensions. 

\subsection*{Acknowledgement}

It is a pleasure to thank Y. Dokshitzer and A. H. Mueller for fruitful
discussions.

\newpage
\appendix
\mysection{Appendix }
\hspace*{\parindent} In this appendix we give explicit expressions  for
the different propagators, for constant fields $\phi_i$ and in the
Landau gauge, both in momentum space and  space-time and for various
limits. For the cutoffs terms $\Delta S_k$ (cf. eq. (\ref{2.10e})) we
use the simplest form

\beq
\label{A.1e}
\Delta S_k(A,B) = {1 \over 2} \int {d^4p \over (2 \pi )^4} \left \{ 
F_{\mu\nu}^n(p) R^A(p, k, \Lambda )
F_{\mu\nu}^n(-p) + B_{\mu\nu}^n(p) R^B(p,k, \Lambda ) 
B_{\mu\nu}^n(-p) \right \}.  \eeq
 
\noi (This form of the cutoff on the fields $A_{\mu}^n$ is consistent 
only in the Landau gauge.) Subsequently
it will be convenient to define the following three 4-index tensors:
 
\bea
\label{A.2e}
&&T_{1, \mu\nu , \rho \sigma}(p) = \delta_{\mu \rho} p_{\nu} 
p_{\sigma} - \delta_{\mu \sigma} p_{\nu}
p_{\rho} - \delta_{\nu \rho} \ p_{\mu} p_{\sigma} + \delta_{\nu 
\sigma} \ p_{\mu} p_{\rho} \ , \nn \\
&&T_{2, \mu\nu , \rho \sigma} = \delta_{\mu \rho} \delta_{\nu 
\sigma} - \delta_{\mu \sigma} \delta_{\nu
\rho} \ , \nn \\
&&T_{3, \mu\nu , \rho \sigma}(p) = T_{2, \mu\nu , \rho \sigma} - {1 
\over p^2} \ T_{1, \mu\nu , \rho \sigma}
(p) \ .
\eea

\noi First we give the three propagators $P^{AA}$, $P^{AB}$ and 
$P^{BB}$ in momentum space:

\bea
\label{A.3e}
&&P_{\mu, \nu}^{AA}(p) = \left ( \delta_{\mu\nu} - {p_{\mu}p_{\nu} 
\over p^2} \right ) P^A(p^2) \ ,\nn \\
&&P^A(p^2) = {\sigma p^2 + 4 (\phi_3 + R^B(p^2)) \over 4p^2 d_1(p^2)} 
\ , \nn \\
&&d_1(p^2) = \sigma p^2\left ( \phi_1 + R^A(p^2) \right ) + 4 \left ( 
\phi_1 + R^A(p^2) \right ) \left ( \phi_3
+ R^B(p^2)\right ) - \phi_2^2 ,
\nn \\ \nn \\
&&P_{\mu, \rho\sigma}^{AB}(p) = i \left ( p_{\rho}\delta_{\mu\sigma} 
- p_{\sigma} \delta_{\mu\rho}\right )
P^{AB}(p^2) \ ,\nn \\
&&P^{AB}(p^2) = {\phi_2 \over  2p^2 d_1(p^2)} \ ,
\nn \\ \nn \\
&&P_{\mu \nu , \rho \sigma}^{BB}(p) =  T_{1, \mu\nu , \rho \sigma} 
\cdot P^{BB}(p^2) + T_{3, \mu\nu , \rho \sigma}
\cdot {1 \over d_2(p^2)}  \ , \nn \\
&&P^{BB}(p^2) = {\phi_1 + R^A(p^2) \over 4p^2 d_1(p^2)} \ , \nn \\
&&d_2(p^2) = h p^2+ 4 R^B(p^2) + 4 \phi_3 \ .
\eea

\noi For later use it is convenient to introduce the propagators 
$P^{FF}$, $P^{FB}$ defined as

\bminiG{A.4e}
\label{A.4ae}
P_{\mu\nu,\rho\sigma}^{FF}(p) = p_{[ \mu} P_{\nu ], [ \rho}^{AA} \ 
p_{\sigma ]} = T_{1, \mu\nu , \rho
\sigma}(p)\ P^A(p^2) \ ,
  \eeeq  \beeq
  \label{A.4be}
P_{\mu\nu,\rho\sigma}^{FB}(p) = p_{[ \mu ,} P_{\nu ], \rho 
\sigma}^{AB} = T_{1, \mu\nu , \rho
\sigma}(p)\ P^{AB}(p^2) \ .
\emini

Many physical phenomena induced by the effective action of our model 
are best understood in terms of the
propagators in ordinary (Euclidean) space-time in the limit where the 
infrared cutoff is removed. For
simplicity we give these propagators also for vanishing UV cutoff 
(which is anyhow non-universal); their
short-distance singularities should be regularized correspondingly. \par

The large distance behaviour of the space-time propagators depends 
often crucially on the combination $\Sigma =
(4\phi_1\phi_3 - \phi_2^2)/\sigma\phi_1$ (defined also in eq. 
(\ref{3.43e})), which vanishes in the
``confining phase'' (cf. section 3). For completeness we give the 
space-time propagators both for finite
$\Sigma$ and for $\Sigma \to 0$. \par

First, for finite $\Sigma$, we have

\bea
\label{A.5e}
P_{\mu, \nu}^{AA}(z) \quad =&& {1 \over 16 \pi^2\sigma \phi_1^2} \left
\{  \delta_{\mu\nu} \left ( {4 \phi_1 \phi_3 \over \Sigma |z|^2} -
{\phi_2^2 \over \sqrt{\Sigma} |z|}  K_1(|z|\sqrt{\Sigma})\right ) 
\right .\nn \\
&&\left . - \partial_{\mu}\partial_{\nu} \left ( {\phi_2^2 \over 
\Sigma^2|z|^2} + {2\phi_1\phi_3 \over \Sigma} \ell n |z| - {\phi_2^2
\over \Sigma^{3/2}|z|} K_1 (|z|\sqrt{\Sigma})\right  ) \right \} \ ,
\nn \\ \nn \\
P_{\mu, \rho\sigma}^{AB}(z) \quad =&& {\phi_2 \over 8
\pi^2\sigma\phi_1}  \left ( \partial_{\rho}\delta_{\mu\sigma} -
\partial_{\sigma}\delta_{\mu\rho}\right ) \left \{ {1 \over 
\Sigma|z|^2} - {1 \over \sqrt{\Sigma}|z|} K_1(|z|\sqrt{\Sigma})\right
\}\ ,
\nn \\ \nn \\
P_{\mu \nu , \rho \sigma}^{BB}(z) \quad =&& {-1 \over 4\pi^2\sigma}
T_{1,  \mu\nu , \rho \sigma}(\partial_z) \left \{ {1 \over \Sigma
|z|^2} - {1 \over \sqrt{\Sigma}|z|}  K_1(|z|\sqrt{\Sigma})\right \} \nn
\\
&&+ {1 \over 16\pi^2\phi_3} T_{1, \mu\nu , \rho \sigma}(\partial_z)
\left \{ {1 \over |z|^2} - {2 \over |z|} \sqrt{{\phi_3 \over h}}
K_1\left (  |z|2\sqrt{{\phi_3 \over h}}\right )\right \} \nn \\ &&+
{\sqrt{\phi_3} \over 2 \pi^2 h^{3/2}} T_{2, \mu \nu ,  \rho\sigma} {1
\over |z|} K_1 \left ( |z|\sqrt{{\phi_3 \over h}} \right )  \ .
\eea

\noi $T_{1, \mu\nu , \rho \sigma}(\partial_z)$ is defined in 
(\ref{A.2e}) with the replacement $p_{\mu}
\to \partial_{\mu}$ and $K_1$ is a Bessel function. The propagators 
$P^{FF}$ and $P^{FB}$ are

\bminiG{A.6e}
\label{A.6ae}
&&P_{\mu \nu, \rho\sigma}^{FF}(z) = {- 1 \over 16 \pi^2\sigma 
\phi_1^2} T_{1, \mu\nu , \rho \sigma}(\partial_z)
\left \{ {4
\phi_1 \phi_3 \over \Sigma |z|^2} - {\phi_2^2 \over \sqrt{\Sigma} 
|z|} K_1(|z|\sqrt{\Sigma})\right \} \ ,
  \eeeq  \beeq
  \label{A.6be}
P_{\mu \nu , \rho\sigma}^{FB}(z) = {\phi_2 \over 8 \pi^2\sigma\phi_1} 
T_{1, \mu\nu , \rho \sigma}(\partial_z)
\left \{  {1 \over \Sigma|z|^2} - {1 \over \sqrt{\Sigma}|z|}
K_1(|z|\sqrt{\Sigma})\right \}\ .
\emini

Finally,  in the limit $\Sigma \to 0$ the space-time propagators 
become (we also replace $\phi_2$ by $2
\sqrt{\phi_1\phi_3}$)

\bea
\label{A.7e}
P_{\mu, \nu}^{AA}(z) \quad =&& {1 \over 16 \pi^2\phi_1} \left \{ 
\delta_{\mu\nu} \left ( {1
\over |z|^2} - {2\phi_3 \over \sigma} \log |z| + \hbox{const.} 
\right ) \right .\nn \\
&&\left . - {1 \over 2} \partial_{\mu}\partial_{\nu} \left ( \log 
|z| - {\phi_3 |z|^2\over 2\sigma} (\log
|z| + \hbox{const.'}) \right ) \right \} \ ,
\nn \\ \nn \\
P_{\mu, \rho\sigma}^{AB}(z) \quad =&& {- \sqrt{\phi_3} \over 8 
\pi^2\sigma\sqrt{\phi_1}} \left ( \partial_{\rho}\delta_{\mu\sigma} - 
\partial_{\sigma}\delta_{\mu\rho}\right ) \log |z| \ ,
\nn \\ \nn \\
P_{\mu \nu , \rho \sigma}^{BB}(z) \quad =&& {1 \over 8\pi^2\sigma} T_{1, 
\mu\nu , \rho \sigma}(\partial_z)
\log |z|\nn \\
&&+ {1 \over 16\pi^2\phi_3} T_{1, \mu\nu , \rho \sigma}(\partial_z) \left
\{ {1 \over |z|^2} - {2 \over |z|} \sqrt{{\phi_3 \over h}} K_1\left ( 
|z|2\sqrt{{\phi_3 \over h}}\right )\right
\} \nn \\
&&+ {\sqrt{\phi_3} \over 2 \pi^2 h^{3/2}} T_{2, \mu \nu , \rho\sigma} 
{1 \over |z|} K_1
\left ( |z|2\sqrt{{\phi_3 \over h}} \right )  \ .
  \eea

\noi The expressions const. and const.$'$ in (\ref{A.7e}) actually 
diverge like $\log(\Sigma)$. A finite result
is obtained, however, for $P^{FF}$ and $P^{FB}$ in the limit $\Sigma \to 0$:

\bminiG{A.8e}
\label{A.8ae}
&&P_{\mu \nu, \rho\sigma}^{FF}(z) = {- 1 \over 16 \pi^2\phi_1} T_{1, 
\mu\nu , \rho \sigma}(\partial_z)
\left \{ {1 \over |z|^2} - {2\phi_3 \over \sigma} \log |z|\right \} \ ,
  \eeeq  \beeq
  \label{A.8be}
P_{\mu \nu , \rho\sigma}^{FB}(z) = {- 1 \over 8 \pi^2\sigma} 
\sqrt{{\phi_3 \over \phi_1}} T_{1, \mu\nu , \rho
\sigma}(\partial_z)   \log |z|\ .
\emini

\end{document}